\documentclass[twocolumn]{aastex701}
\usepackage{amsmath}
\usepackage{natbib}
\graphicspath{{figures/}}

\newcommand{\mbh}{M_{\rm BH}}
\newcommand{\msun}{M_{\odot}}
\newcommand{\abh}{a_{\rm BH}}
\newcommand{\rg}{r_{\rm g}}
\newcommand{\s}{{\rm s}}
\newcommand{\ye}{Y_{\rm e}}
\newcommand{\cm}{{\rm cm}}

\newcommand{\Ps}{P_{\rm S}}
\newcommand{\Pw}{P_{\rm W}}
\newcommand{\Ts}{T}
\newcommand{\Bz}{B_{\rm 0}}

\begin{document}
\shortauthors{D. Issa \& O. Gottlieb}
\shorttitle{ Can Direct-Horizon Collapsars Explain Early $r$-Process Enrichment?}
\title{Neutron-Rich Disks, Nickel-Rich Ejecta:\\
Can Direct-Horizon Collapsars Explain Early $r$-Process Enrichment?}

\author[orcid=0009-0005-2478-7631,sname='Danat Issa']{Danat Issa}
\email[show]{danat@mit.edu}
\affiliation{Department of Physics and Kavli Institute for Astrophysics and Space Research, Massachusetts Institute of Technology, Cambridge, MA 02139, USA}
\affiliation{Center for Interdisciplinary Exploration and Research in Astrophysics (CIERA), Physics and Astronomy, Northwestern University, Evanston, IL 60202, USA}

\author[0000-0003-3115-2456]{Ore Gottlieb}
\email{oregottlieb@gmail.com}
\affiliation{Department of Physics and Kavli Institute for Astrophysics and Space Research, Massachusetts Institute of Technology, Cambridge, MA 02139, USA}

\begin{abstract}
Very massive stars above the pair-instability mass gap may collapse directly into black holes surrounded by massive disks. These ``direct-horizon collapsars'' accrete at $\dot{M} \gtrsim 10\,M_\odot\,{\rm s}^{-1}$ and have been proposed to eject a substantial amount of $r$-process material, powering ``super-kilonovae''. We present the first three-dimensional general-relativistic magnetohydrodynamic simulations of direct-horizon collapsars with M1 neutrino transport, following a $\approx 150\,M_\odot$ helium core from collapse through the escape of magnetically driven outflows. Extreme accretion rates neutronize the disks to $Y_{\rm e} \approx 0.1$--$0.25$ and, at early times, allow them to satisfy the gravitational-instability and cooling criteria for fragmentation. However, the strong poloidal fields needed to eject the neutron-rich gas promptly also drain the disks and lift their electron degeneracy. The neutron-rich ejecta composition is set in a brief phase, such that the ejecta carry $\sim 0.1$--$0.3\,M_\odot$ of nuclei with $A \geq 69$, including $\lesssim 2\times10^{-4}\,M_\odot$ of lanthanides, alongside $\sim 1$--$4\,M_\odot$ of $^{56}$Ni. Assuming proportional retention and mixing of Eu and Fe, these yields imply $[{\rm Eu/Fe}]\approx-0.4$ to $+0.3$, and simple dilution estimates permit enrichment at $[{\rm Fe/H}]\lesssim-2.6$. Direct-horizon collapsars are therefore potential prompt contributors to iron and $r$-process enrichment at low metallicity. Their predicted Eu/Fe ratios fall below those of strongly $r$-process-enhanced stars, but uncertainties in the relative yields, particularly of iron, may close this gap. The ${}^{56}{\rm Ni}$-rich ejecta would power luminous supernovae with rise times of weeks, potentially accompanied by exceptionally powerful gamma-ray bursts, rather than lanthanide-rich super-kilonovae.
\end{abstract}

\section{Introduction} \label{sec:introduction}

The rapid neutron capture process ($r$-process) is responsible for about half of the abundance of the heavy elements beyond iron in the Universe. It proceeds through successive neutron captures onto seed nuclei on timescales shorter than $\beta$-decay, requiring environments with extreme neutron richness \citep[][]{Arnould2007,Cowan2021,Thielemann2026}. The subsequent radioactive decay of these unstable, neutron-rich nuclei powers electromagnetic transients known as kilonovae, providing an observable signature of $r$-process nucleosynthesis \citep{Li1998,Metzger2010,Metzger2020}.

The first multimessenger detection of a binary neutron star (BNS) merger, GW170817 \citep{Abbott2017,Nakar2020,Margutti2019}, accompanied by a kilonova detection, provided compelling evidence that BNS mergers are a major site of $r$-process nucleosynthesis \citep[e.g.,][]{Kasen2017,Pian2017,Tanvir2017}. While the inferred rates and expected yields from BNS mergers are broadly consistent with the bulk of the solar $r$-process inventory, it remains unclear whether they are the sole $r$-process factories \citep[e.g.,][]{Hotokezaka2018,Cote2019}.

The presence of $r$-process elements, such as europium, in extremely metal-poor halo stars \citep{Sneden2008,Frebel2015,Frebel2026} and in the ultra-faint dwarf galaxy Reticulum II \citep{Ji2016_nature,Ji2016} indicates that $r$-process enrichment must have occurred at early cosmic times. However, the canonical delay-time distribution of BNS mergers suggests that a significant fraction of systems merge only after tens to hundreds of Myr \citep[e.g.,][]{Dominik2012,Vigna-Gomez2018}. This temporal distribution can be difficult to reconcile with the prompt enrichment inferred for the most metal-poor systems. A fast BNS merger channel has been proposed to address this tension \citep{Beniamini2019,Safarzadeh2019,Beniamini2024,Maoz2025}. However, the rates and yields of fast-merger populations, when constrained by gravitational-wave, gamma-ray burst, and elemental abundance data, remain in tension with the observed $r$-process enrichment history \citep[e.g.,][]{Saleem2025}. This suggests that additional sources capable of generating similarly neutron-rich ejecta may be required to enrich low-metallicity environments \citep[e.g.,][]{Wehmeyer2015,Haynes2019,vandeVoort2020,Soker2026}.

Massive stars, which evolve and explode on Myr timescales, naturally provide such prompt enrichment channels.\footnote{Other prompt or short-delay $r$-process channels associated with massive stars and their compact remnants have also been proposed, including magnetar giant flares \citep{patel_direct_2025} and common-envelope supernovae \citep{Grichener2019,Grichener2025,Soker2025}. However, their nucleosynthetic yields remain uncertain.} However, neutrino-driven core-collapse supernovae are not expected to robustly synthesize the heavy $r$-process nuclei as intense neutrino irradiation from the proto-neutron star (PNS) drives the ejecta proton-rich \citep{Qian1996,Hudepohl2010,Arcones2013,Wanajo2013}. Nevertheless, subsets of rapidly rotating, strongly magnetized explosions can give rise to magnetically driven collimated fast outflows with rapid expansion and reduced neutrino exposure, conditions that are more favorable for $r$-process nucleosynthesis.

One proposed channel is magnetorotational supernovae, in which magnetic stresses tap the rotational energy of a rapidly rotating PNS, thereby ejecting neutron-rich material \citep{Winteler2012,Nishimura2015}. However, numerical models find mostly weak $r$-process, with more substantial production of the heaviest nuclei favored by exceptionally strong, suitably configured magnetic fields \citep{Halevi2018,Mosta2018,Reichert2021,Reichert2023,Zha2024}. Complementary local calculations of proto-magnetar outflows explore magnetic acceleration \citep{Metzger2007,Metzger2008}, magnetic confinement followed by high-entropy ejection \citep{Prasanna2024,Prasanna2025}, and disk--magnetosphere interactions in accreting systems \citep{Prasanna2026}. These calculations find some favorable conditions for heavy $r$-process nucleosynthesis, but prescribe the central remnant and its neutrino emission rather than deriving them from stellar collapse, and neglect interactions with the surrounding stellar material. Whether the assumed conditions arise and persist, and how much $r$-process material ultimately escapes, therefore requires assessment in global calculations.

Another proposed channel is collapsars \citep{Woosley1993,macfadyen_collapsars_1999}, in which the PNS collapses to form a spinning black hole (BH) surrounded by an accretion disk. If the disk reaches sufficiently high densities and becomes mildly electron-degenerate, electron and positron captures can drive the gas neutron-rich \citep{Popham1999,Chen2007}, enabling $r$-process nucleosynthesis. As magnetic flux accumulates on and threads the BH, magnetically driven, collimated disk outflows are launched \citep{Tchekhovskoy2011}, carrying neutron-rich material into the surrounding medium and enriching the interstellar medium. The resulting disk winds have been proposed as prolific sites of $r$-process nucleosynthesis \citep{Siegel2019,Brauer2021}.

However, efficient $r$-process nucleosynthesis requires the neutrino cooling timescale to be shorter than the accretion timescale, so that the disk material can cool and neutronize before being swallowed by the BH. Additionally, unbinding this neutron-rich gas demands strong magnetic stresses capable of launching powerful outflows. These requirements are partially in tension: strong magnetic fields enhance angular-momentum transport, thereby shortening the accretion timescale, while also increasing thermal and magnetic pressure, which thicken the disk and reduce its cooling efficiency \citep{Issa2025}. As a result, the conditions that favor neutronization tend to oppose those required to unbind the neutron-rich material. \citet{Gottlieb2026} showed that only engines with unusually high accretion rates may eject heavy $r$-process material. In such systems, the disk can sustain neutron-rich conditions while still powering magnetically driven outflows, potentially enabling neutron-rich outflow production.

Recent observational evidence increasingly supports the existence of such very massive stars above the pair-instability mass gap, ``direct-horizon collapsars''. In particular, the detection of high-mass BH mergers such as GW231123 \citep[e.g.,][]{LVK2025_GW231123,gottlieb_spinning_2025,Popa2025}, together with stellar evolution models and observations of stars with inferred masses $M \gtrsim 200\,M_\odot$ in the LMC \citep{dekoter:97,crowther:10,crowther:16,brands:22}, including 30 Doradus \citep{schneider:18science,schneider:18b}, suggests that extremely massive progenitors can form in nature even at metallicities in the LMC ($\sim 0.5 Z_{\odot}$). These systems provide a physically motivated extension of ordinary collapsars and offer a potential channel for the prompt $r$-process enrichment of the early Universe for two reasons.

First, their disks can sustain extremely high accretion rates, $\dot{M} \gtrsim 10\,M_\odot\,\s^{-1}$ \citep{gottlieb_spinning_2025}, allowing them to remain dense and efficiently neutrino-cooled while strongly magnetized. Under such conditions, the disk may become neutron-rich and simultaneously power magnetically driven outflows, satisfying the requirements for both neutronization and mass ejection. Second, because the earliest generations of stars formed in metal-poor environments, reduced line-driven mass loss would have enabled massive stars to retain larger helium cores \citep{Vink2001,Vink2005,Heger2003} and greater angular momentum \citep{Yoon2005}, potentially increasing the fraction of direct-horizon collapsars at high redshift when early enrichment occurred. Direct-horizon collapsars thus represent promising candidates for prompt $r$-process production that operates on the short timescales required to enrich extremely metal-poor halo stars.

Semi-analytic models suggest that these stars are indeed conducive to neutronization and may power super-kilonovae \citep{siegel_super-kilonovae_2022,Hernandez-Morales2026}. However, as in the case of ordinary collapsars, the nonlinear coupling between magnetic stresses, neutrino cooling, and accretion dynamics can only be captured by self-consistent simulations that include magnetic fields, neutrino transport, and the collapsing star. Each previous simulation lacked at least one of them. Those that included the collapsing star were either axisymmetric without magnetic fields \citep{Uchida2019,Shibata2026} or three-dimensional without neutrinos \citep{gottlieb_spinning_2025}. \citet{agarwal2026} performed 3D general-relativistic magnetohydrodynamic (GRMHD) simulations of tori around $80$--$3000\,\msun$ BHs, including weak interactions but not the star, and inferred the outflow composition from that of the disk.

In this \emph{Letter}, we present the first three-dimensional GRMHD simulations of direct-horizon collapsars that incorporate a two-moment (M1) neutrino transport scheme. We find that their collapsar disks can become neutron-rich owing to extreme mass accretion rates, $\dot{M} \gtrsim 10\, \msun\,\rm{s}^{-1}$, and can launch powerful outflows in the presence of very strong magnetic fields. However, the magnetic flux that launches the outflows also drains the disk and lifts its degeneracy within less than a second, such that only $\approx 10^{-2}\,\msun$ maintains $\ye\lesssim 0.25$. We outline our simulation pipeline, from progenitor evolution, core-collapse, and BH formation to the subsequent launch of outflows, in Section~\ref{sec:methods}, present our simulation results in Section~\ref{sec:results}, and conclude and discuss their broad implications in Section~\ref{sec:conclusions}.

\section{Numerical setup} \label{sec:methods}

\subsection{Core collapse to BH formation}

Our progenitor star\footnote{Publicly available at \url{https://zenodo.org/records/3406357}.} begins with a bare helium core of mass $ M_{\rm He,0} = 250.75\,\msun $ and metallicity $ Z = 0.001 \approx 0.07 Z_\odot$, assuming that the hydrogen-rich envelope has been stripped away. Owing to stellar wind-driven, helium-rich mass loss, the helium-core mass is reduced to $M_{\rm He,cc}\approx 149\,\msun$ by the onset of core collapse \citep{Renzo2020b}.

In modeling the core collapse to BH formation, we neglect the gas at densities $\rho\leq 2\times 10^3\,{\rm g\,cm^{-3}}$ in the outer part of the core and evolve the inner $ M_{\rm sim} \approx 140\,\msun $ of the star, enclosed within a radius of $ R_{\rm He} \approx 5.6\times 10^9\,{\rm cm} $. We simulate the collapse using the general-relativistic hydrodynamic code \texttt{GR1D}\footnote{Available at \url{https://github.com/evanoconnor/GR1D}.} \citep{OConnor2010}, which solves the general relativistic hydrodynamics equations in spherical coordinates, using the radial gauge and polar-slicing metric \citep{Romero_1996}, and employs a microphysical equation of state \citep[LS220;][] {Lattimer_LS220}. We adopt a 1D radial grid with 1800 points, with uniform spacing from the origin out to $2\times 10^{6}\,\rm{cm}$ and logarithmic spacing beyond that to $1.34\,R_{\rm He} = 7.5\times 10^{9}\,\rm{cm}$.

We prescribe the rotation of the star directly at the core collapse phase in \texttt{GR1D} using the parameters that are favorable for $r$-process material formation from \citet{siegel_super-kilonovae_2022},
\begin{equation}
\begin{split}
    l(r,\theta) &= f_{\rm K} l_{\rm K} (r) \sin^2\theta \cdot
    \begin{cases}
        \left(\frac{r}{r_{\rm b}}\right)^p, \quad r < r_{\rm b}, \\
        1, \quad r_{\rm b} \le r \le R_{\rm He},
    \end{cases}
\end{split}
\end{equation}
where $l_{\rm K}(r) = \sqrt{GM(<r)r}$ is the specific Keplerian angular momentum with enclosed mass, $M(<r)$; we choose $f_{\rm K} = 0.2$, $r_{\rm b} = 1.5\times 10^9\,\cm$ and $p=4.5$.\footnote{We explored $f_{\rm K} \in \{0.1, 0.2\}$ together with $r_{\rm b} \in \{ 1.5\times 10^8, 1.5\times 10^9 \}\,\rm{cm}$; the chosen set of parameters results in the largest, most massive disk with a smaller BH, conducive to neutron-rich disk formation.} The gas at $r<r_{\rm b}$ would fall directly onto the BH, whereas exterior shells have sufficient angular momentum to circularize outside the innermost stable circular orbit once they fall in.

\texttt{GR1D} cannot evolve through the formation of a spacetime singularity, and it terminates when the lapse, $\alpha \equiv 1/\sqrt{-g^{tt}}$, drops close to zero. For our purposes, we define the BH formation when the central lapse falls below $\alpha \approx 10^{-2}$ \citep[see e.g.,][]{oconnor_open-source_2015}. We integrate the gravitational mass and the total angular momentum enclosed within the region where $\alpha\leq 10^{-2}$, which results in the BH mass, $\mbh = 23.9\,\msun$, and dimensionless spin, $\abh = 0.36$. The time of BH formation and the initial BH mass and spin depend on the initial prescribed angular momentum profile. For comparison, \citet{gottlieb_spinning_2025} collapsed the same helium core without prescribing a rotational profile at the core-collapse stage, which resulted in an initial BH mass of $\approx 40\,\msun$.

\subsection{Post-BH formation}\label{sec:postbh}

To model the post-BH formation phase, we map the mass density, gas pressure, radial and azimuthal velocity profiles of the star at the time of BH formation from \texttt{GR1D} into the 3D GRMHD code \texttt{$\nu$H-AMR} \citep{hamr_paper, Issa2025}. \texttt{$\nu$H-AMR} is a GPU-accelerated code that evolves the neutrino-GRMHD equations and employs a relativistic Helmholtz equation of state \citep[EoS;][]{timmes_accuracy_2000} and two-moment (M1) neutrino transport. We summarize the neutrino transport implementation in Appendix \ref{appendix:nu_transport}, and refer to \citet{Issa2025} for the full description. 

In $\nu$\texttt{H-AMR}, the spacetime metric is a stationary Kerr metric with fixed BH mass and spin parameters, $\mbh$ and $\abh$, which does not include the contribution of self-gravity of the shells of the star. 
Mapping pressure and velocity profiles without modification would result in a violent relaxation of the gas towards its new equilibrium, and lead to an unphysical amount of unbound ejecta early on.
To mitigate the relaxation, we rescale the hydrodynamic quantities using $m(r) = \mbh / M_{\rm enc} (r) \leq 1$, where $M_{\rm enc} (r)$ is the enclosed gravitational mass, including the BH mass, interior to the radial coordinate $r$. 
We modify the gas pressure by solving 
\begin{equation}
\begin{split}
    \quad P_{\rm g}^{\texttt{H-AMR}}(R_{\rm He}) &= m(R_{\rm He})\, P_{\rm g}^{\texttt{GR1D}} (R_{\rm He}), \\
    \frac{\partial P_{\rm g}^{\rm \texttt{H-AMR}} (r)}{\partial r} &= m(r)\,\frac{\partial P_{\rm g}^{\rm \texttt{GR1D}} (r)}{\partial r}.
\end{split}
\end{equation}
The radial and azimuthal velocities are rescaled by $\sqrt{m(r)}$. The rescaling of the pressure preserves hydrostatic equilibrium in the potential of the BH, and the rescaling of the azimuthal velocity preserves the circularization radius of each shell, $r_{\rm circ} = l^2 m/(G\mbh) = l^2/(G M_{\rm enc})$. We map the density directly from \texttt{GR1D} into \texttt{$\nu$H-AMR}, and set the initial electron fraction, $\ye=0.5$. Since \texttt{GR1D} uses LS220 EoS, we do not remap the temperature. Instead, we solve for the temperature using the Helmholtz EoS from the remapped density and pressure profiles. This remapping allows us to preserve the hydrostatic pressure balance in the star.

The magnetic field is instrumental in driving outflows from the inner parts of the accretion disk. We prescribe a range of initial magnetic field configurations to explore their effects on outflow production and composition: strong and weak poloidal fields, $\Ps$ and $\Pw$, a toroidal field, $\Ts$, and an unmagnetized model, $\Bz$ (see Table~\ref{tab:models}). In the strong poloidal field model, $\Ps$, we initialize the field through the azimuthal component of the covariant magnetic vector potential,
\begin{equation}\label{eq:Aphi_Bs}
\begin{split}
    A_{\varphi} &= A_0 \sin^2\theta \, \max\!\left[ f_{\rm s}(r) - f_{\rm s}(r_{\rm hole}),\, 0 \right], \\
    f_{\rm s}(r) &\equiv \frac{r^2}{r^2 + R_{\rm core}^2} - \frac{r^3}{R_{\rm s}^3},
\end{split}
\end{equation}
where $R_{\rm core} = 10^{8}\,{\rm cm}$ is the core radius, $R_{\rm s} = 5\times 10^{9}\,{\rm cm}$ is the radius beyond which the field vanishes ($R_{\rm s} \leq R_{\rm He}$), $r_{\rm hole} = 6\,\rg \approx 0.2\,R_{\rm core}$ is the radius inside which the field vanishes, and $A_0$ is the normalization constant (see below). Subtracting $f_{\rm s}(r_{\rm hole})$ makes $A_\varphi$ continuous at the inner cutoff. The magnetic field in the coordinate frame is obtained through $B^{i} = (-g)^{-1/2}\,[ijk]\,\partial_j A_k$, where $[ijk]$ is the antisymmetric symbol. We define the central field strength $B_{\rm c} \equiv 2A_0/R_{\rm core}^2$, the strength of the uniform field, aligned with the stellar rotation axis, that Eq.~\eqref{eq:Aphi_Bs} would describe at $r \ll R_{\rm core}$ without the hole.\footnote{Since $r_{\rm hole} \approx 0.2\,R_{\rm core}$, the field in the core is not uniform: on the equator it decreases from $\approx 0.9\,B_{\rm c}$ at $r_{\rm hole}$ to $\approx 0.25\,B_{\rm c}$ at $R_{\rm core}$.} For $R_{\rm core} \ll r \ll R_{\rm s}$, the field falls off as $B \propto r^{-2}$. In the model with a weaker poloidal magnetic field, $\Pw$, we instead adopt a dipolar profile,
\begin{equation}\label{eq:Aphi_Bw}
\begin{split}
    A_{\varphi} &= A_0 \sin^2\theta \, \max\!\left[ f_{\rm w}(x) - f_{\rm w}(R_{\rm s}),\, 0 \right], \\
    f_{\rm w}(r) &\equiv \frac{R_{\rm core}\, r^2}{r^3 + R_{\rm core}^3}, \quad x \equiv \sqrt{r^2 - r_{\rm hole}^2},
\end{split}
\end{equation}
for $r \geq r_{\rm hole}$, and $A_\varphi = 0$ otherwise. Since $f_{\rm w}(R_{\rm s}) \approx R_{\rm core}/R_{\rm s}$, the field vanishes inside $r \approx 0.26\,R_{\rm core} \approx 7\,\rg$. The equatorial field is $\approx B_{\rm c} \equiv 2A_0/R_{\rm core}^2$ out to $r \approx 0.3\,R_{\rm core}$ and decreases to $\approx 0.15\,B_{\rm c}$ at $R_{\rm core}$. For $R_{\rm core} \ll r \ll R_{\rm s}$, the field approaches a dipole with magnetic moment $A_0 R_{\rm core}$. In the model with toroidal field, $\Ts$, we set $A_{\theta}$ to the profile of Eq.~\eqref{eq:Aphi_Bs}, which gives a purely toroidal field with the same equatorial profile as in $\Ps$, so $B_{\rm c}$ has the same meaning. In all magnetized models, $A_0$ is set by the ratio of the maximum pressure to the maximum magnetic pressure in the star, $\beta = p_{\max}/(b^2_{\max}/2)$. Table~\ref{tab:models} lists the resulting $B_{\rm c}$.

We use spherical polar coordinates that are uniform in $(\log r, \theta, \varphi)$, where $(r, \theta, \varphi)$ are Kerr–Schild coordinates. The radial grid extends from $ r_{\rm in} = 0.9\,r_{\rm h} = 1.7\,\rg $ (where $r_{\rm h} = (1 + \sqrt{1-\abh^2})\,\rg$ is the event horizon radius) out to $ r_{\rm out} = 10^5\,\rg \approx 63\,R_{\rm He} $, where $\rg = G M_{\rm BH}/c^2$ is the BH gravitational radius. The simulations use a base grid of $N_{\rm r} \times N_\theta \times N_{\varphi} = 448\times 288\times 112$ cells, and we use one level of static mesh refinement (SMR) in the $\varphi$ direction within $\theta \in (\pi/6, 5\pi/6)$, which gradually increases the $\varphi$ resolution from the pole to the equator \citep{hamr_paper}. In model $\Ps$, we apply 1 SMR layer for all cells within $r \in (7, 100)\,\rg$, $\theta \in (\pi/3, 2\pi/3)$ and $\varphi \in (0, 2\pi)$, resulting in the effective resolution in the disk $N_{\rm r} \times N_\theta \times N_{\varphi} = 896\times 576\times 448$. To adequately resolve the magnetorotational instability (MRI; \citealt{mri_paper}) in models with weaker initial poloidal field, $\Pw$, and purely toroidal field, $\Ts$, we apply 2 SMR layers covering the same region, which brings the effective resolution in the disk to $N_{\rm r} \times N_\theta \times N_{\varphi} = 1792\times 1152\times 896$. We verify that the MRI is sufficiently resolved in the poloidal field simulations $\Ps$ and $\Pw$. In the toroidal field simulation, $\Ts$, the MRI is resolved after $t \gtrsim 0.15\,\rm{s}$. We show the grid structure of model $\Pw$ and the MRI criterion profiles in Appendix~\ref{appendix:mri_criterion} (Fig.~\ref{fig:MRI_disk}).

\begin{table}[ht]
\centering
\caption{A summary of the model parameters: strong poloidal field, $\Ps$, weak poloidal field, $\Pw$, toroidal field, $\Ts$, and the unmagnetized model, $\Bz$. Columns show the initial central field strength, $B_{\rm c} = 2A_0/R_{\rm core}^2$ (Section~\ref{sec:postbh}), the initial magnetic vector potential, the effective resolution in the disk, and the final simulation time.}
\label{tab:models}
\renewcommand{\arraystretch}{1.2}
\small
\setlength{\tabcolsep}{3pt}
\begin{tabular}{lcccc}
    \hline
    Model & $B_{\rm c}$ [$10^{14}\,\rm{G}$] & $A_i$ & $N_r, N_\theta, N_\varphi$ & $t_{\max}$ [s] \\
    \hline
    $\Ps$ & $0.8$ & $A_\varphi$, Eq.~\eqref{eq:Aphi_Bs} & $896,576,448$ & $4.26$ \\
    $\Pw$ & $0.4$ & $A_\varphi$, Eq.~\eqref{eq:Aphi_Bw} & $1792,1152,896$ & $4.29$ \\
    $\Ts$ & $1.5$ & $A_\theta$, Eq.~\eqref{eq:Aphi_Bs} & $1792,1152,896$ & $3.56$ \\
    $\Bz$ & --- & --- & $448,288,112$ & $6.04$ \\
    \hline
\end{tabular}

\end{table}

\begin{figure}
    \centering
    \includegraphics[width=\columnwidth]{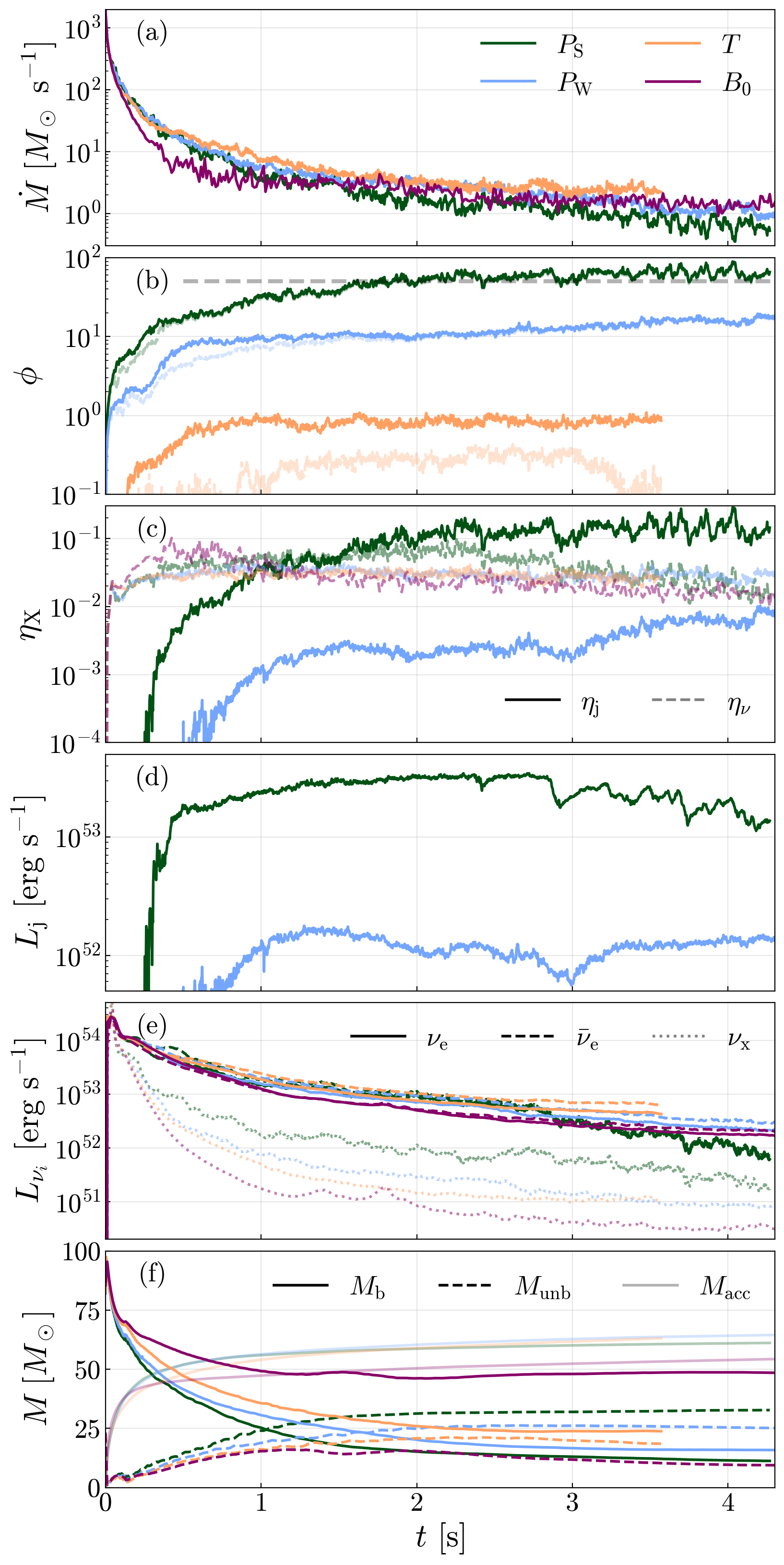}
    \caption{Time evolution of the quantities near the BH. Panel (a): BH mass accretion rate begins at $\dot{M} \sim 10^3 \msun\,\rm{s}^{-1}$, and falls roughly as $\propto t^{-1.2}$. Panel (b): Normalized magnetic flux on the BH (solid lines) and signed magnetic flux threading BH's northern hemisphere (faint lines) show that the MAD state ($\phi = 50$, dashed gray horizontal line) is reached only in model $\Ps$. Panel (c): Jet efficiency (solid lines) exceeds $10^{-4}$ at the time of launching in models $\Ps$ and $\Pw$; the neutrino radiative efficiency (dashed lines) remains below $0.1$ in all models. Panel (d): Jet power exceeds $10^{53}\,\rm{erg\,s}^{-1}$ in $\Ps$ and reaches $\sim 10^{52}\,\rm{erg\,s}^{-1}$ in $\Pw$. Panel (e): Luminosities of electron neutrinos (solid), electron antineutrinos (dashed), and heavy-lepton neutrinos (dotted) at $r = 100\,\rg$ exceed $10^{54}\,\rm{erg\,s}^{-1}$ at early times and decline with $\dot{M}$. Panel (f): Bound ($M_{\rm b}$, solid) and unbound ($M_{\rm unb}$, dashed) mass, and accreted mass ($M_{\rm acc}$, faint lines).}
    \label{fig:BH_time_evolution_plot}
\end{figure}

\section{Numerical results} \label{sec:results}

We first present the accretion onto the BH and the launching of outflows (Section~\ref{sec:engine}), then the disk composition (Section~\ref{sec:disk}) and fragmentation (Section~\ref{sec:disk_fragmentation}), and finish with the ejecta (Section~\ref{sec:ejecta}) and their nucleosynthesis (Section~\ref{sec:tracers}).

\subsection{Accretion and outflow launching} \label{sec:engine}

Fig.~\ref{fig:BH_time_evolution_plot} depicts the time evolution of the quantities near the BH. Fig.~\ref{fig:BH_time_evolution_plot}(a) shows the mass accretion rate,
\begin{equation} \label{eq:mdot}
    \dot{M} = - \iint_{r=5\rg} \sqrt{-g} \rho u^r d\theta d\varphi,
\end{equation}
where $\sqrt{-g}$ is the volume factor, $\rho$ is the rest mass density in the comoving frame and $u^r$ is the radial component of the 4-velocity. Owing to the massive compact progenitor, the mass accretion rate reaches $\dot{M} \approx 10^{3}\,\msun\,\rm{s}^{-1}$ at BH formation. Subsequently, the mass accretion rate declines as $\dot{M} \propto t^{-1}$--$t^{-1.3}$, depending on the model, faster than in typical collapsar simulations \citep[e.g.,][]{gottlieb_black_2022}. Model $\Bz$, which has no magnetic field, accretes at a lower rate at $t \approx 0.15$--$0.8\,\rm{s}$, due to inefficient angular momentum transport in the absence of MRI \citep{mri_paper}; at later times, its $\dot{M}$ is comparable to that of the magnetized models.

Fig.~\ref{fig:BH_time_evolution_plot}(b) presents the magnetically arrested disk \citep[MAD;][]{Tchekhovskoy2011} parameter (dark lines),
\begin{equation} \label{eq:phiBH}
    \phi = \frac{\sqrt{4\pi}}{2\sqrt{\dot{M}}} \iint_{r=r_{\rm h}} \sqrt{-g} |B^r| d\theta d\varphi,
\end{equation}
where $B^r$ is the radial component of the magnetic field vector measured in the coordinate frame. Horizontal dashed gray line marks the MAD saturation value, $\phi \approx 50$ \citep{Tchekhovskoy2011}. In the model with a strong initial poloidal magnetic field, $\Ps$, the MAD state is achieved at $t\approx 1.5\,\rm{s}$. In the models with weaker initial fields, namely $\Pw$, $\Ts$, and $\Bz$, the accretion flow does not reach the MAD state during the simulation. Fig.~\ref{fig:BH_time_evolution_plot}(b) additionally shows the signed magnetic flux threading the BH northern hemisphere (faint lines),
\begin{equation} \label{eq:phiBH_N}
    \phi_{\rm N} = \frac{\sqrt{4\pi}}{\sqrt{\dot{M}}} \iint_{r=r_{\rm h}} \sqrt{-g} B^r d\theta d\varphi\, \Big|_{\theta \leq \pi/2}.
\end{equation}
% s
In models with initial poloidal fields, $\Ps$ and $\Pw$, $\phi_{\rm N} \approx \phi$. On the other hand, in the model with an initial toroidal field, $\Ts$, the magnetic flux on the BH is considerably less coherent, such that $\phi_{\rm N}$ is smaller than $\phi$ by a factor of a few, potentially too weak to launch BH-driven outflows within the short accretion time in direct-horizon collapsars \citep[see e.g.,][]{Chan2026}.

Fig.~\ref{fig:BH_time_evolution_plot}(c) illustrates the magnetized jet and neutrino efficiencies,
\begin{equation} \label{eq:eta_jet}
    \eta_{\rm j} = \frac{L_{\rm j}}{\dot{M}c^2}, \quad \eta_{\nu} = \frac{L_{\nu}}{\dot{M}c^2},
\end{equation}
where $L_{\rm j}$ and $L_\nu$ are the jet and total neutrino luminosities.
The jet launching, $\eta_{\rm j} \gtrsim 10^{-4}$, occurs once $\phi \gtrsim 10$ in model $\Ps$, and at $\phi \sim 3$--$5$ in model $\Pw$. In the model with an initial toroidal field, $\Ts$, the jet does not launch due to an inefficient MRI dynamo, which fails to generate sufficiently strong poloidal magnetic flux by the end of the simulation, $t \approx 3.6\,\rm{s}$.
Neutrino radiative efficiencies remain $\approx 2$--$5\%$ for most of the evolution and below $10 \%$ in all models, similar to expectations for rapidly accreting disks \citep{Chen2007}. Only in model $\Ps$ does $\eta_{\rm j}$ eventually exceed $\eta_{\nu}$, as it approaches the MAD state. In model $\Ps$, $\eta_{\nu}$ increases at $t \approx 2$--$3\,\rm{s}$, because the disk becomes less dense and more transparent, so fewer neutrinos are trapped and advected onto the BH \citep{DiMatteo2002,Chen2007}. It decreases again at $t \gtrsim 3\,\rm{s}$, when the disk mass drops below $0.1\,\msun$ (Section~\ref{sec:disk}).

The jet power, shown in Fig.~\ref{fig:BH_time_evolution_plot}(d), is
\begin{equation} \label{eq:l_jet}
    L_{\rm j} = - \iint_{r=5\rg} \sqrt{-g} (T^r_t + \rho u^r) d\theta d\varphi \,\Big|_{\sigma>1},
\end{equation}
which is computed by integrating the radial energy flux, $T^r_t$, excluding the rest mass contribution and only accounting for material with magnetization $\sigma\equiv b^2/\rho>1$. The variation in initial field configuration leads to a variety of jet powers, exceeding $10^{53}\,\rm{erg\,s^{-1}}$ in $\Ps$ and reaching $\sim 10^{52}\,\rm{erg\,s^{-1}}$ in $\Pw$. 

Fig.~\ref{fig:BH_time_evolution_plot}(e) depicts the neutrino luminosities of each species evolved in the simulation (electron neutrinos, $\nu_{\rm e}$, electron anti-neutrinos, $\bar{\nu}_{\rm e}$, and heavy-lepton neutrinos, $\nu_{\rm x}$),
\begin{equation} \label{eq:l_nu}
    L_{\nu_i} = - \iint_{r=100\rg} \sqrt{-g} R^r_{\nu_i; t} d\theta d\varphi,
\end{equation}
where the neutrino energy flux, $R^{r}_{\nu_i; t}$ is defined in Eq.~\eqref{eq:stress-energy-tensor}, at $r=100\,\rg$, near the outer edge of the disk.
The neutrino luminosities exceed $10^{54}\,\rm{erg\,s^{-1}}$ at early times, when $\dot{M} \gtrsim 10^2\,\msun\,\rm{s}^{-1}$, and roughly follow $\dot{M}$ as it decreases. These neutrino fluxes irradiate the outflows and alter their composition (see Section~\ref{sec:tracers}).

\subsection{Disk composition} \label{sec:disk}

\begin{figure*}
    \centering
    \includegraphics[width=\linewidth]{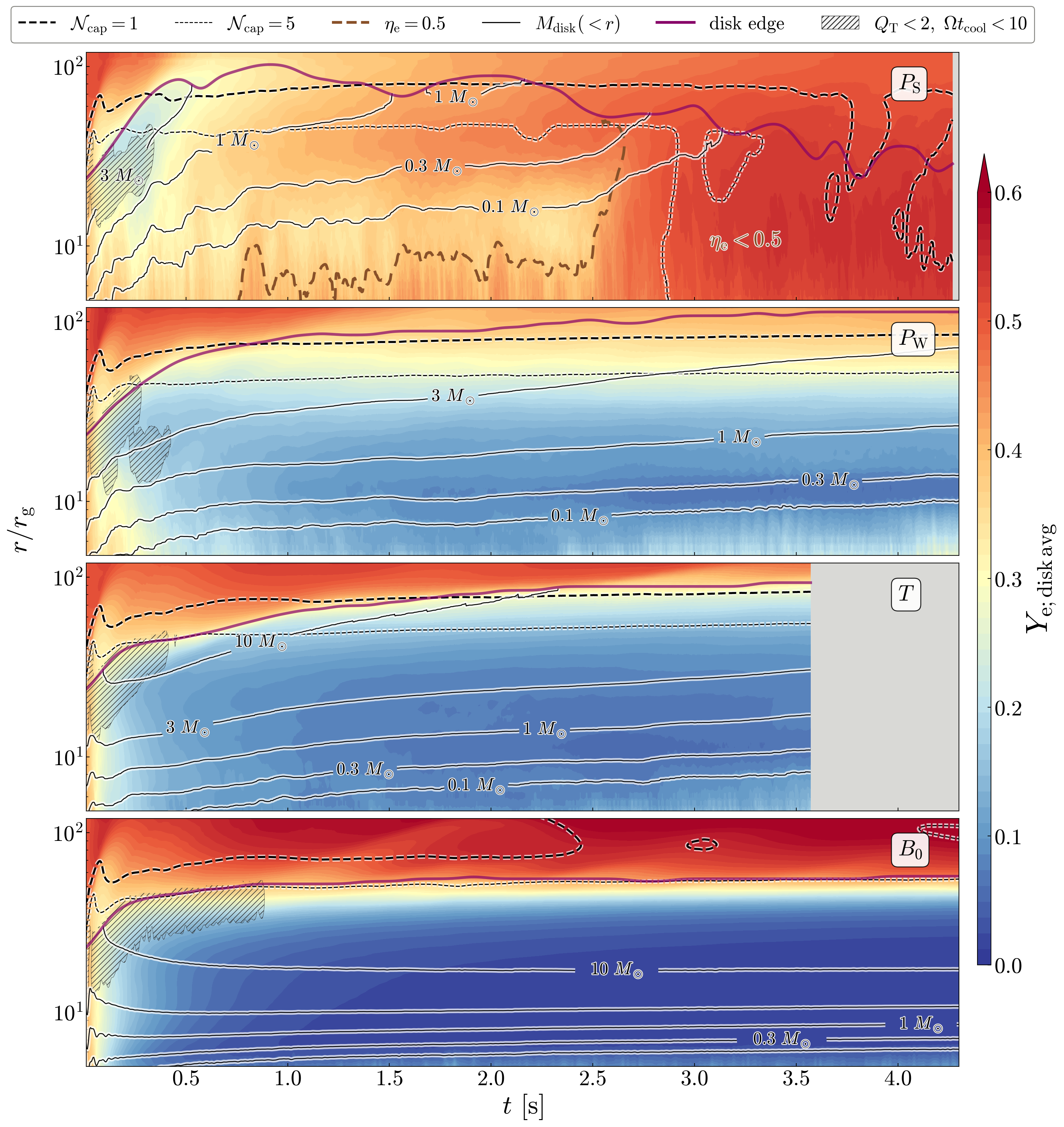}
    \caption{Density-weighted average $\ye$ in the disk as a function of time and radius [Eq.~\eqref{eq:ye_rt_avg}] shows the average composition of the disk (neutron-rich in blue, neutron-poor in red). The disk in the model with initially strong poloidal field, $\Ps$, reaches $\ye \approx 0.25$ only at early times, $t \lesssim 0.5\,\rm{s}$, whereas in models $\Pw$, $\Ts$, and $\Bz$ it remains neutron-rich throughout the simulation. In models $\Pw$, $\Ts$, and $\Bz$, the radial extent of the neutron-rich disk correlates with the ``ignition'' region, where the number of electron and positron captures per nucleon within the local residence time, $\mathcal{N}_{\rm cap}$, exceeds unity (thick dashed black lines mark $\mathcal{N}_{\rm cap}=1$; thin dashed black lines mark $\mathcal{N}_{\rm cap}=5$). Brown dashed line encloses the non-degenerate region, $\eta_{\rm e} < 0.5$, in model $\Ps$. Solid purple lines show the disk edge, and solid black lines show the contours of the enclosed disk mass, $M_{\rm disk}(<r)$. Hatched regions mark where the disk is both gravitationally unstable and rapidly cooling, $Q_{\rm T} < 2$ and $\Omega t_{\rm cool} < 10$ (see Section~\ref{sec:disk_fragmentation}). Gray areas mark times without data.}
    \label{fig:Ye_spacetime_diagram}
\end{figure*}

Fig.~\ref{fig:Ye_spacetime_diagram} depicts the density-weighted average electron fraction in the disk,
\begin{equation} \label{eq:ye_rt_avg}
    Y_{\rm e;\,disk\,avg} = \frac{\langle \rho \ye \rangle_{\rm d}}{\langle \rho \rangle_{\rm d}},
\end{equation}
where we compute the averages within the disk defined as
\begin{equation} \label{eq:disk_avg}
    \langle X \rangle_{\rm d} = \frac{\int_{\pi/2 - \theta_{\rm d}}^{\pi/2 + \theta_{\rm d}} d\theta \int d\varphi \sqrt{-g} X}{\int_{\pi/2 - \theta_{\rm d}}^{\pi/2 + \theta_{\rm d}} d\theta \int d\varphi \sqrt{-g}},
\end{equation}
with the density-weighted angular scale height of the disk,
\begin{equation} \label{eq:disk_scale}
    \theta_{\rm d} = \frac{\iint \sqrt{-g} \rho |\theta - \pi/2| d\theta d\varphi}{\iint \sqrt{-g} \rho d\theta d\varphi},
\end{equation}
computed at each radius and time.
In the model with the strong initial poloidal field, $\Ps$, the disk reaches $Y_{\rm e;\,disk\,avg} \approx 0.25$ only in the beginning, $t \lesssim 0.5\,\rm{s}$. In all other models, the disk remains neutron-rich, $Y_{\rm e;\,disk\,avg} \approx 0.1$, throughout the simulation.

We define the disk edge (solid purple line) as the radius where the local specific angular momentum drops below $0.75\,l_{\rm K}$. In model $\Ps$, the disk gradually shrinks after $t \approx 1\,\rm{s}$, whereas it spreads outwards in the other magnetized models, $\Pw$ and $\Ts$, or remains unchanged in the absence of angular momentum transport in model $\Bz$. The disk in model $\Ps$ has an enclosed mass of $3\,\msun$ (solid black lines), but quickly loses its mass to BH accretion and outflows, and after $t > 3\,\rm{s}$ its mass drops below $0.1\,\msun$. In model $\Pw$, the disk loses its mass more gradually, together with its neutron-rich inner part. The disk in model $\Ts$, with an initial toroidal field, is a factor of $\approx 2$--$3$ more massive than in $\Pw$ within the same radius. In model $\Bz$, without magnetic fields, the inner part becomes denser and more massive over time.

To identify what sets the disk composition, we estimate how many electron and positron captures each nucleon undergoes while it resides in the disk,
\begin{equation}
    \mathcal{N}_{\rm cap} = \lambda_{eN}\,\min(t, \bar{t}_{\rm in}),
\end{equation}
where $\lambda_{eN}$ is the total capture rate per nucleon. The residence time is the local inflow time,
$\bar{t}_{\rm in} = r/\bar{v}_{\rm in}$,
where
\begin{equation}
    \bar{v}_{\rm in} = -\frac{\iint \sqrt{-g}\,\rho u^r\,d\theta\,d\varphi}{\iint \sqrt{-g}\,\rho u^t\,d\theta\,d\varphi}
\end{equation}
is the mass-weighted net inflow velocity of the disk. In the outer disk, where $\bar{t}_{\rm in} > t$, we cap the residence time at $t$, since no gas can have been exposed to captures for longer than the age of the disk. We refer to the region where $\mathcal{N}_{\rm cap} \geq 1$ as the ``ignition'' region. There, captures can reset $\ye$ toward its local equilibrium value, such that the disk composition is set by the local conditions rather than inherited from the infalling stellar gas.

In all models, the ignition region covers the bulk of the disk (bounded by the thick dashed black line). Raising the threshold to $\mathcal{N}_{\rm cap} \geq 5$ (thin dashed black line) leaves this result unchanged at least until $t \approx 3\,{\rm s}$, after which the disk $\ye$ in model $\Ps$ is set by only $1$--$5$ captures. Since $\ye$ in this model begins to rise at $t \approx 0.5\,{\rm s}$, well before $\mathcal{N}_{\rm cap}$ drops below 5, the rise cannot be attributed to inefficient captures. Therefore, the local equilibrium electron fraction itself must be increasing. Two effects can raise it: the lifting of the electron degeneracy and neutrino absorptions.

In the transparent regime, which neglects neutrino absorptions, the equilibrium electron fraction, $Y_{\rm e;\,eq,\,tr}$, of hot accretion disks, $T\gtrsim 1.3\,\rm MeV$, is low when the electrons are degenerate \citep{Beloborodov2003}, and $Y_{\rm e;\,eq,\,tr}$ rises towards $\approx 0.5$ as the degeneracy is lifted toward $\eta_{\rm e} < 0.5$ (dashed brown line in Fig.~\ref{fig:Ye_spacetime_diagram}). Model $\Ps$ is the only one in which the inner disk becomes non-degenerate at $t\gtrsim 0.75\,\rm{s}$, and the entire disk at $t\gtrsim 2.5\,\rm{s}$. The disks in other models remain degenerate, $\eta_{\rm e} \ge 0.5$ everywhere in the plotted range, and, therefore, neutron-rich, consistent with the self-regulation of the degeneracy in neutrino-cooled disks \citep{SiegelMetzger2018,Siegel2019,agarwal2026}.

\begin{figure*}[hbtp]
    \centering
    \includegraphics[width=0.98\linewidth]{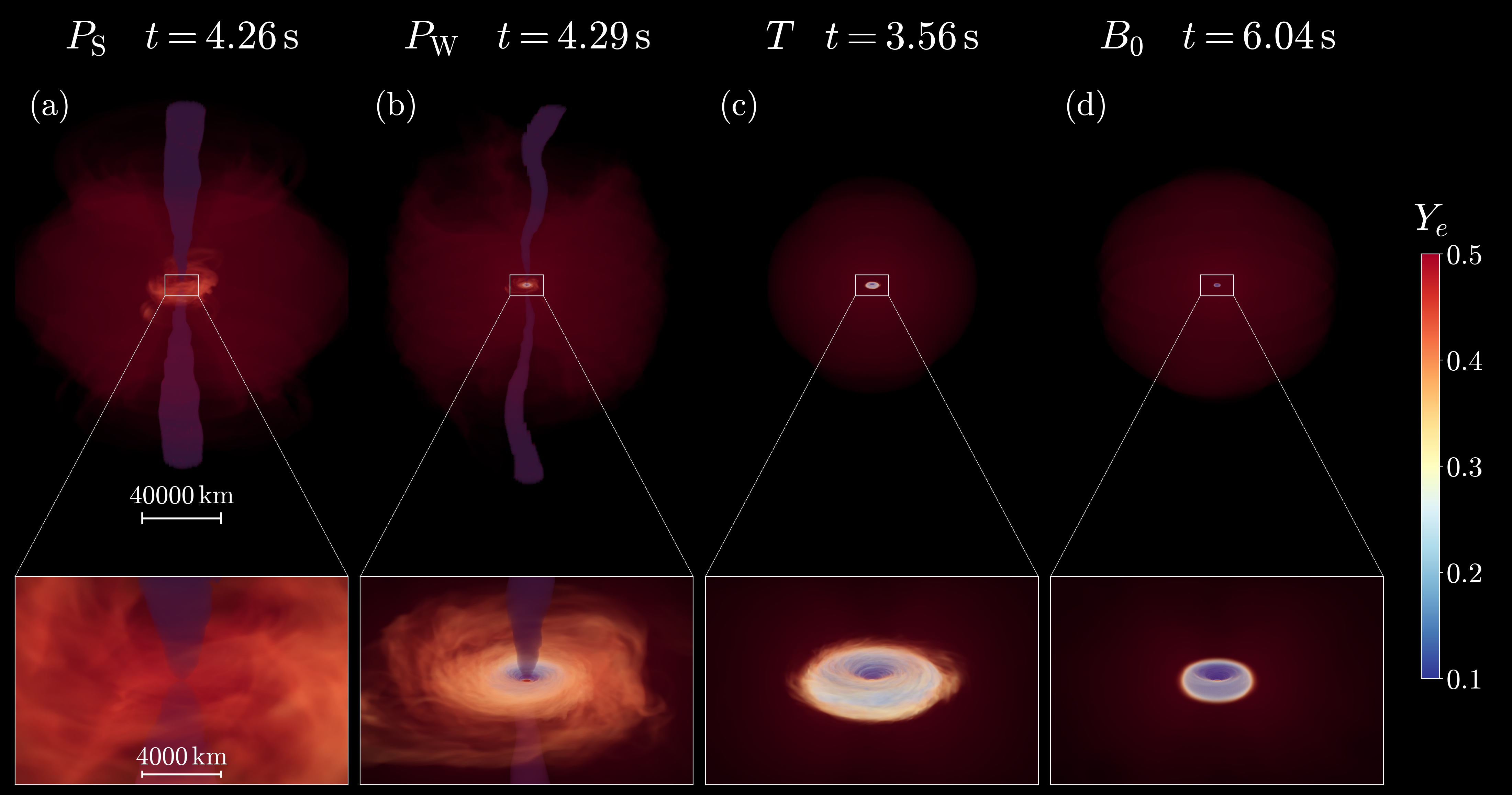}
    \caption{3-dimensional rendering of $\ye$ in all models at the times indicated above each column. The top row shows the outflows on large scales, and the bottom row zooms in on the accretion disk (neutron-rich in blue, neutron-poor in red); jet ($\mathcal{B}>1$) is shown in purple. Panel (a): outflows in the model with strong poloidal fields, $\Ps$, launch early and remain powerful; by $t = 4.26\,\rm{s}$, the disk has largely lost its neutron-rich material. At large distances away from the BH, the outflows mix with the infalling neutron-poor gas. Panel (b): outflows in the model with weaker poloidal fields, $\Pw$, launch later, are weaker, and unbind the neutron-poor material from the disk surface, while the inner disk remains neutron-rich. Panels (c) and (d): In models $\Ts$ and $\Bz$, the disk remains neutron-rich, but no outflows are launched to unbind the neutron-rich gas.}
    \label{fig:Ye_3dmaps}
\end{figure*}

In dense disks, neutrino absorptions drive $\ye$ above the transparent equilibrium value.
To estimate this shift, we compute the equilibrium including absorptions from the net lepton-number source term, $d\ye/dt |_{Y_{\rm e;\,eq,\,abs}} = 0$, using the local $\nu_{\rm e}$, $\bar \nu_{\rm e}$ fields.
At early times, when disks are still opaque, $Y_{\rm e;\,eq,\,abs} - Y_{\rm e;\,eq,\,tr} \lesssim 0.05$, which drops to $\lesssim 0.02$ as the disks become more transparent later.
Therefore, the disk protonizes mainly because its degeneracy is lifted, which raises $Y_{\rm e;\,eq,\,tr}$, and absorptions drive $\ye$ further up. They continue to raise $\ye$ in the outflows, which are launched from the less dense disk surface, where the equilibrium $\ye$ is higher than in the midplane \citep{Miller_2020}.
As a result, the disk outflows that launch at late times carry material that is no longer neutron-rich.

Thus, the extreme accretion rate is necessary but not sufficient for neutron-rich ejecta. At $t \approx 0.5$--$2\,\rm{s}$, the three magnetized models accrete at rates within a factor of a few of each other, $\dot{M} \approx 2$--$20\,\msun\,\rm{s}^{-1}$ [Fig.~\ref{fig:BH_time_evolution_plot}(a)], yet only in model $\Ps$, which approaches the MAD state, does the disk drain, lose its degeneracy, and protonize. The strong magnetic flux needed to eject the neutron-rich gas also removes the conditions that produce it, as in ordinary collapsars \citep{Issa2025}. The extreme $\dot{M}$ of direct-horizon collapsars relieves this tension only briefly, opening a window at $t \lesssim 0.5\,\rm{s}$ in $\Ps$, during which the outflows are launched while the disk is still neutron-rich.

\subsection{Disk fragmentation} \label{sec:disk_fragmentation}

The hatched regions in Fig.~\ref{fig:Ye_spacetime_diagram} mark where the disk is both gravitationally unstable and cools rapidly through electron and positron captures on free nucleons,
\begin{equation} \label{eq:frag}
    Q_{\rm T} = \frac{c_{\rm s} \langle\Omega\rangle_{\rm d}}{\pi G \Sigma_{\rm d}} < 2, \quad
    \langle\Omega\rangle_{\rm d}\, t_{\rm cool} = \langle\Omega\rangle_{\rm d} \frac{\langle u_{\rm g} \rangle_{\rm d}}{\langle \dot Q_{\nu} \rangle_{\rm d}} < 10.
\end{equation}
Here $c_{\rm s} = (\langle p \rangle_{\rm d}/\langle \rho \rangle_{\rm d})^{1/2}$ is the approximate isothermal sound speed, $\Omega=u^\varphi/u^t$ is the angular velocity, $u_{\rm g}$ is the internal energy density, $\dot Q_{\nu}$ is the net (emission minus absorption) neutrino cooling rate, and $\langle \cdot \rangle_{\rm d}$ denotes the disk average [Eq.~\eqref{eq:disk_avg}]. $\Sigma_{\rm d}$ is the $\varphi$-averaged rest-mass surface density of the disk,
\begin{equation}
    \Sigma_{\rm d} = \frac{1}{2\pi r} \int d\varphi \int_{\pi/2-\theta_{\rm d}}^{\pi/2+\theta_{\rm d}} d\theta \sqrt{-g} \rho u^t.
\end{equation}
The threshold values follow estimates in \citet{ChenMetzger2025}, who find disk fragmentation for $\Omega t_{\rm cool} \lesssim 5$--$10$ at $Q_{\rm T} \approx 1$. We use $Q_{\rm T} < 2$ so as to include marginally unstable disks.
All magnetized models meet both criteria at early times, $t \lesssim 0.25$--$0.45\,\rm{s}$, at $r \approx 10$--$45\,\rg$, while $\dot M \gtrsim 10\,\msun\,{\rm s^{-1}}$ [Fig.~\ref{fig:BH_time_evolution_plot}(a)]. The disks in $\Ts$ and $\Bz$ remain at $Q_{\rm T} < 2$, but their neutrino cooling is too slow at later times. In $\Bz$, the marginally unstable region resides near the outer disk edge until $t \approx 0.9\,\rm{s}$.  We discuss the implications of this estimate in Section~\ref{sec:fragmentation}.

\subsection{Unbound ejecta} \label{sec:ejecta}

Fig.~\ref{fig:BH_time_evolution_plot}(f) shows the accreted mass (faint solid lines), unbound (dashed lines) and bound (solid lines) masses, defined by the Bernoulli parameter cutoff,
\begin{equation} \label{eq:bernoulli_mhd}
    \mathcal{B} = -1 - u_t \left(1 + \frac{u_{\rm g} + P_{\rm g} + b^2}{\rho}\right)>0.
\end{equation}
The BH mass roughly doubles by $t \approx 0.1\,\rm{s}$ and more than triples by $t \approx 1\,\rm{s}$, and the total unbound and bound masses roughly plateau after $t \gtrsim 2\,\rm{s}$. At $t \lesssim 1\,\rm{s}$, the unbound masses are comparable in all models, $M_{\rm unb}\approx 15$--$20\,\msun$, part of which results from the initial relaxation of the mapped star on the fixed Kerr metric (Section~\ref{sec:postbh}). At later times, the initial magnetic field configuration sets the final unbound mass. In models $\Ps$, $\Pw$, and $\Ts$, $M_{\rm unb}\approx 30, 25$, and $20\,\msun$, respectively. In the unmagnetized model, $\Bz$, accreted and unbound masses are smaller due to inefficient angular momentum transport and the absence of magnetic pressure. The unbound mass peaks at $\approx 17\,\msun$ at $t \approx 1.2\,\rm{s}$ and decreases to $\approx 12\,\msun$ by the end of the simulation, as part of the marginally unbound gas becomes bound again. Because the BH grows rapidly while the metric is held fixed, these unbound masses are likely overestimated (Section~\ref{sec:caveats}).

Fig.~\ref{fig:Ye_3dmaps} illustrates $\ye$ of the outflows in all models at the times indicated in the figure. In model $\Ps$ [Fig.~\ref{fig:Ye_3dmaps}(a)], the outflows launch early, while the disk is still neutron-rich, and carry part of this neutron-rich material away from the disk. At large distances from the BH, they mix with the infalling neutron-poor stellar gas and appear predominantly neutron-poor (shown in red), and by this time the disk itself has largely lost its neutron-rich material. In contrast, in the model with a weaker initial poloidal field, $\Pw$ [Fig.~\ref{fig:Ye_3dmaps}(b)], the disk midplane remains neutron-rich, but the delayed, weaker outflows predominantly unbind the material from the disk surface, where it is already neutron-poor. Fig.~\ref{fig:Ye_3dmaps}(c) shows that model $\Ts$ is unable to launch sufficiently strong disk outflows, and the disk remains neutron-rich. The disk in model $\Bz$ [Fig.~\ref{fig:Ye_3dmaps}(d)] is more compact and neutron-rich. Overall, neutron-rich outflows reach large radii only if they are launched while the disk is still degenerate, which requires either a strong initial poloidal field or an MRI dynamo that quickly generates strong poloidal fields.

\subsection{Nucleosynthesis in the outflows} \label{sec:tracers}

\begin{figure}
    \centering
    \includegraphics[width=\columnwidth]{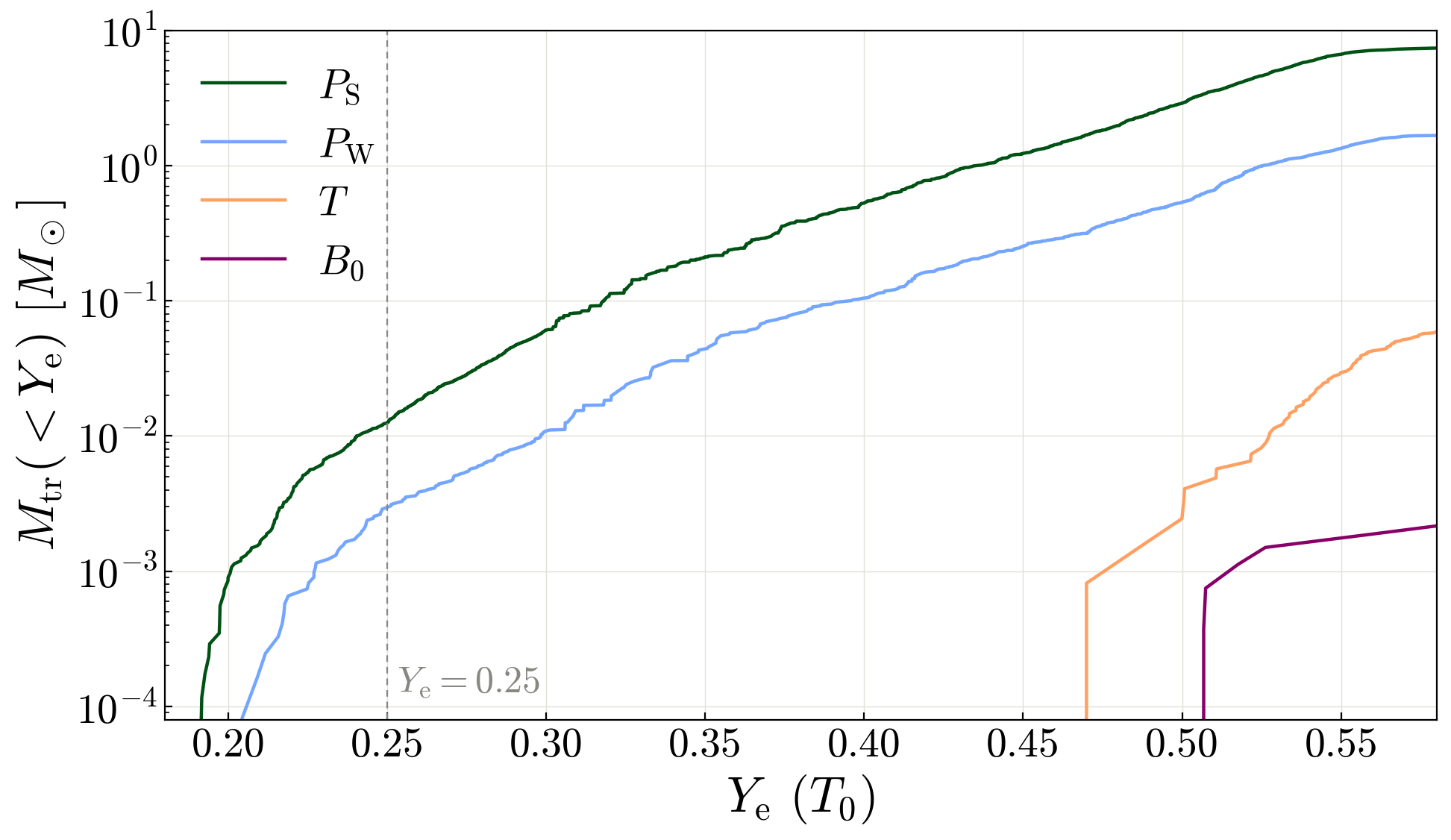}
    \caption{Cumulative mass of the tracers with $T_{\rm tr;\,peak}\geq5\,$GK as a function of $\ye$, excluding the shock-heated stellar gas, at the time the tracers last cool through $T_0 = \min(T_{\rm tr;\,peak}, 8\,{\rm GK})$, i.e., at the start of the \texttt{SkyNet} calculation. Only the poloidal field models, $\Ps$ and $\Pw$, have tracers with $\ye < 0.25$ (vertical dashed line), whereas models without jets, $\Ts$ and $\Bz$, eject only material with $\ye \gtrsim 0.45$.}
    \label{fig:Ye_tracers}
\end{figure}

To estimate the yields in the disk ejecta, and in the absence of tracer particles in \texttt{$\nu$H-AMR}, we reconstruct Lagrangian trajectories from the data in post-processing. We sample $\mathcal{N}_{\rm T}=3\times 10^5$ tracers from the unbound ($\mathcal{B}>0$), outflowing ($u^r>0$) gas, excluding gas heated by a shock during the relaxation of the initial conditions (see Appendix~\ref{appendix:tracers_comp}). We integrate the tracers backward in time, recording $\rho$, $T$, $\ye$ and the neutrino radiation field along each trajectory, and re-integrate $\ye$ along them with corrected neutrino absorption rates (Appendix~\ref{appendix:tracers_ye}). Those are post-processed in the nuclear reaction network \texttt{SkyNet} \citep{Lippuner2017}, starting in nuclear statistical equilibrium (NSE) when each tracer cools through $T_0 = \min(T_{\rm tr;\, peak}, 8\,\rm{GK})$, where $T_{\rm tr;\, peak}\geq 5\,\rm{GK}$ is the tracer's peak temperature, and follow the nucleosynthesis to $t=10^9\,$s (Appendix~\ref{appendix:tracers_nuc}). Fig.~\ref{fig:Ye_tracers} shows the $\ye$ mass distribution of the tracers at the start of their \texttt{SkyNet} trajectories. Only models $\Ps$ and $\Pw$ eject gas with $\ye < 0.25$, amounting to $0.013$ and $0.003\,\msun$, respectively.

\begin{figure}
    \centering
    \includegraphics[width=\columnwidth]{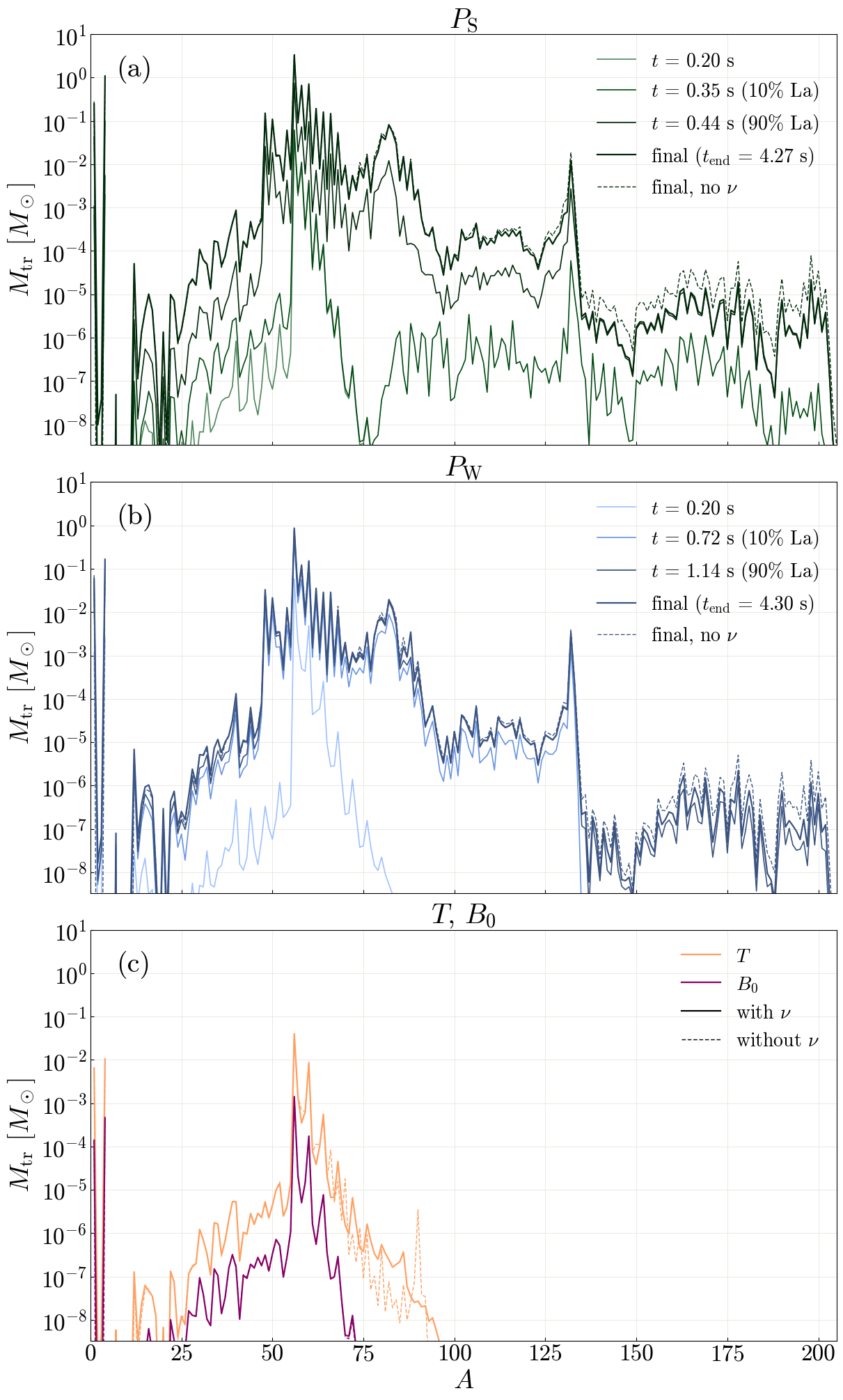}
    \caption{Nucleosynthesis yields of the traced unbound ejecta, excluding the shock-heated stellar gas, as a function of mass number, $M_{\rm tr}(A)$, computed with \texttt{SkyNet}, in models $\Ps$ (a), $\Pw$ (b), and $\Ts$ and $\Bz$ (c). In panels (a) and (b), thin lines show the yields of the tracers that cooled through $T_0$ by the indicated times, including those by which $10\%$ and $90\%$ of the final lanthanide mass is produced, and thick solid lines show the final yields. Dashed lines show the final yields computed without neutrino absorptions after network initialization. Only models $\Ps$ and $\Pw$ produce nuclei beyond the first $r$-process peak, $A \approx 80$, including the second, $A \approx 130$, and third, $A \approx 195$, peaks.}
    \label{fig:yields}
\end{figure}

Fig.~\ref{fig:yields} shows the yields in all models. The estimated total mass of the $r$-process ejecta \citep{Arnould2007} is $M_{\rm r-p} = M_{\rm ej} (A \ge 69) \approx 0.3\,\msun$ in $\Ps$ and $M_{\rm r-p} \approx 0.1\,\msun$ in $\Pw$, with most of the $r$-process ejecta cooling through $T_0$ in the first $\sim 1\,\rm{s}$ in both models.
In model $\Ps$ ($\Pw$), lanthanides ($57 \leq Z \leq 71$) amount to $M_{\rm La} \approx 1.4\times 10^{-4}\,\left(1.1\times 10^{-5}\right)\,\msun$, $90\%$ of which is produced by the ejecta that cools through $T_0$ by $t \approx 0.44\,(1.14)\,\rm{s}$. They constitute $\approx 1\%\,(0.4\%)$ of the ejecta mass below $\ye=0.25$, the threshold for lanthanide production \citep{Lippuner2015}. No model produces actinides.

In Fig.~\ref{fig:yields}, the yield pattern peaks at $A = 56$, most of which is ${}^{56}\rm Fe$. Its total mass is $M(A=56) \approx 3.6$ and $0.75\,\msun$ in models $\Ps$ and $\Pw$, and only $\approx 0.04$ and $0.002\,\msun$ in $\Ts$ and $\Bz$, and is almost entirely produced by tracers with $\ye \gtrsim 0.49$, indicating that it was ${}^{56}\rm Ni$ that eventually decayed to ${}^{56}\rm Fe$. Since all of the tracers included in the nucleosynthesis here reach peak temperatures $T_{\rm tr;\,peak} \ge 5\,\rm{GK}$ \citep[complete Si burning;][]{Woosley1973}, we likely underestimate the relative yields of lighter elements from discarding low-$T$ tracers (Appendix~\ref{appendix:tracers_nuc}).
The ejecta mass is likely overestimated because we neglect the BH mass growth and self-gravity, so these masses should be regarded as order-of-magnitude estimates subject to the limitations discussed in Section~\ref{sec:caveats}.

We run each tracer twice: with $\nu_e$ and $\bar\nu_e$ absorption driven by the neutrino field recorded along its trajectory, and without neutrino absorption. Appendix~\ref{appendix:tracers} describes the tracer sampling, integration, and nucleosynthesis calculations. Fig.~\ref{fig:tracer_trajectories} (Appendix~\ref{appendix:tracers_traj}) shows that the low-$\ye$ tracers spend the first $\sim 2$\,s near the BH and the disk, where the neutrino fluxes are highest. As a result, neutrino absorption raises their $\ye$ \citep[cf.][]{Miller_2020}, reducing the yields beyond $A \sim 130$ by a factor of $\sim 2$--$3$ relative to the calculation without absorption (dashed lines in Fig.~\ref{fig:yields}).

\section{Discussion and conclusions}
\label{sec:conclusions}

We have presented the first 3D GRMHD simulations of direct-horizon collapsars with M1 neutrino transport, initialized from the collapse of an $\approx150\,\msun$ helium core into a BH with natal $\mbh=23.9\,\msun$ and $\abh=0.36$. We followed the subsequent $>3\,{\rm s}$ of accretion with four initial magnetic field configurations. Our simulations reveal a tension between disk neutronization and the ejection of neutron-rich gas. Extreme accretion rates allow the disks to neutronize, while strong poloidal magnetic flux enables jets and neutron-rich outflows. The jet-launching models produce $\gtrsim10^{-1}\,\msun$ of $r$-process nuclei, including $\sim10^{-5}$--$10^{-4}\,\msun$ of lanthanides. However, the strong magnetic flux also rapidly depletes the disk through accretion and outflows, lifting its electron degeneracy and driving it toward higher $\ye$. The neutron-rich gas is set during a brief early phase, and the ejecta in the jet-launching models are dominated by iron-group material, including $\sim1$--$4\,\msun$ of ${}^{56}\mathrm{Ni}$. Before exploring the astrophysical implications, we discuss the principal uncertainties affecting the disk evolution, ejecta masses, and relative nucleosynthetic yields.

\subsection{Limitations and uncertainties} \label{sec:caveats}

\begin{enumerate}
    \item The spacetime is a stationary Kerr metric with fixed $\mbh$ and $\abh$, and the self-gravity of the star is neglected, yet the BH mass more than triples within $t \approx 1\,\rm{s}$. The deeper potential of the growing BH would bind part of the unbound ejecta in our simulations. For example, in the weak-field limit, the Bernoulli parameter of gas, $\mathcal{B}$, at $r \gtrsim 200\,\rg$ would decrease by $\sim G \Delta M_{\rm BH} / (rc^2) \sim 10^{-2}$, comparable to that of most of the unbound ejecta (see Appendix~\ref{appendix:tracers_comp} and Fig.~\ref{fig:ejecta_ye_B}). The BH growth also alters the disk structure and neutrino field. Since more massive BHs lead to higher $\ye$, the ejecta would likely be less neutron-rich had this growth been included.

    \item Our simulations do not include $\alpha$-particle recombination. Free nucleons that recombine into $\alpha$ particles in the expanding outflows release $\approx 7\,\rm{MeV}$ per nucleon, raising $\mathcal{B}$ by $\approx 7.5\times10^{-3}X_\alpha$, where $X_\alpha$ is the alpha-particle mass fraction, and $r$-process heating adds to it in the neutron-rich ejecta. These effects partly offset the BH growth in the bulk of the ejecta at $\ye \approx 0.5$. However, they may not be able to offset it in the neutron-rich ejecta. The reason is that the $\ye < 0.25$ gas resides at $r \approx 100$--$600\,r_{\rm g}$ at the end of the simulations (Fig.~\ref{fig:tracer_trajectories}), where the growth of the BH to $M_{\rm BH} \gtrsim 85\,M_\odot$ lowers $\mathcal{B}$ by $\approx 0.004$--$0.03$. Since $X_\alpha \leq 2\ye \lesssim 0.5$, recombination and $r$-process heating together raise $\mathcal{B}$ by only $\lesssim 7\times10^{-3}$, such that our heavy $r$-process and lanthanide yields should be considered approximated upper limits.

    \item Our results rely on a single progenitor model, whose rotation profile, magnetic field, and composition at core collapse are uncertain. We prescribe the angular momentum profile at the onset of collapse, choosing the parameters that result in the largest, most massive disk with a smaller BH, and the initial BH mass and spin depend on this choice. Furthermore, the stellar magnetic field is not constrained by stellar evolution models, and our models show that it determines the outflow mass and composition.
\end{enumerate}

With these limitations in mind, we consider the implications for BHs in the pair-instability supernova mass gap (Section~\ref{sec:final_bh}), disk fragmentation and possible NS formation (Section~\ref{sec:fragmentation}), the resulting electromagnetic transients (Section~\ref{sec:transients}), $r$-process production (Section~\ref{sec:rprocess}), and its enrichment at low metallicities (Section~\ref{sec:enrichment}).

\subsection{Final BH mass and spin} \label{sec:final_bh}

\citet{gottlieb_spinning_2025} proposed direct-horizon collapsars as a formation channel for the massive BH binaries observed by LIGO--Virgo--KAGRA, such as GW231123. By the end of our simulations, $\mbh \gtrsim 85\,\msun$ in the magnetized models, and $\approx 10$--$25\,\msun$ of bound gas remains on the grid. If this gas is eventually accreted, together with the marginally unbound outflows that BH growth would rebind (Section~\ref{sec:caveats}), the final BH masses are $\sim 100$--$130\,\msun$. BHs that grow rapidly by accreting weakly magnetized gas should spin up to nearly maximal spins. However, magnetic stresses reduce the angular momentum carried into the BH, and jets extract its rotational energy \citep{issa_spinevol_2026,issa_collapsar_2026}. Since the magnetic flux in our models becomes dynamically important (near-MAD state) only after the BH has already grown substantially, the BHs would likely remain at moderate-to-high spins. During the early, weakly magnetized accretion the BHs spin up, and only model $\Ps$ reaches the MAD state ($t\approx1.5\,\rm{s}$; Section~\ref{sec:engine}), by which time $\mbh$ has more than tripled. Accreting its remaining bound gas in the MAD state could lower the spin of the BH in $\Ps$ by a few tenths \citep{issa_spinevol_2026}, whereas the BHs in $\Pw$, $\Ts$, and $\Bz$ would remain rapidly spinning. Hence, the BHs in the latter models would resemble the components of GW231123, $m_1 = 137^{+23}_{-18}\,\msun$ and $m_2 = 101^{+22}_{-50}\,\msun$, with spins $\chi_1 = 0.90^{+0.10}_{-0.19}$ and $\chi_2 = 0.80^{+0.20}_{-0.51}$ \citep{LVK2025_GW231123}. We leave the feedback of the evolving BH mass and spin on the disk, the star, and the outflows for future work.

\subsection{Collapsar disk fragmentation}
\label{sec:fragmentation}

Collapsar disks accreting at $\dot M \gtrsim 10\,\msun\,{\rm s^{-1}}$ can become gravitationally unstable \citep{Toomre1964}, and, if neutrino cooling is fast enough, they may fragment \citep{Piro2007}. \citet{Metzger2024} proposed this mechanism as a formation channel for subsolar-mass neutron stars (NSs) in neutron-rich disks.
At $t \lesssim 0.45\,{\rm s}$, the disks are both gravitationally unstable and rapidly cooling at $r \approx 10$--$45\,\rg$, where $\ye \approx 0.2$--$0.35$. For $\Sigma\propto r^{-1}$ and $\Sigma\simeq2\rho H$, the characteristic clump mass is the Jeans mass \citep{ChenMetzger2025},
\begin{equation}
    M_J\sim\rho H^3
    \simeq\frac{\Sigma H^2}{2}
    \simeq\left(\frac{H}{r}\right)^2
    \frac{M_{\rm disk}(<r)}{4\pi},
\end{equation}
such that $M_J \sim 0.003$--$0.07\,\msun$ for $M_{\rm disk}(<r) \sim 1$--$10\,\msun$ and $H/r \sim 0.2$--$0.3$. This is below the cold Chandrasekhar mass at this composition, $M_{\rm Ch}\simeq1.45\,\msun\,(\ye/0.5)^2\approx0.2$--$0.7\,\msun$. Therefore, forming subsolar NSs would require the clumps to grow through accretion and mergers, or to neutronize further as they contract \citep{Metzger2024,ChenMetzger2025,Lerner2026}. At later times, neutrino cooling is too slow for fragmentation in all models, and the unstable regions are more likely regulated by gravitoturbulence that suppresses the formation of bound fragments \citep{Gammie2001}, although $\alpha$-particle dissociation, which we omit, would favor collapse. Since our simulations also omit self-gravity, and $Q_{\rm T}$ is only an approximate diagnostic, whether clumps form at all remains to be tested. The simultaneous requirements of a massive disk and rapid cooling may favor these extreme progenitors, as comparable conditions could be harder to achieve in ordinary collapsars.

\subsection{Electromagnetic signatures} \label{sec:transients}

In the jet-launching models, the traced ejecta contain $\approx0.75$--$3.6\,\msun$ of ${}^{56}\mathrm{Ni}$, substantially exceeding those of typical core-collapse SNe \citep{Anderson2019}. In the strongly magnetized model, the yield lies within the range predicted for pair-instability SNe \citep{HegerWoosley2002,Kasen2011} and is comparable to the nickel mass inferred for SN~2007bi under a radioactive-powering interpretation \citep{GalYam2009}. These yields remain uncertain and may be overestimated (Section~\ref{sec:caveats}). Nevertheless, even a reduced yield of $\sim1\,\msun$ of ${}^{56}\mathrm{Ni}$ would power a luminous transient with a peak bolometric luminosity of a few $\times10^{43}\,{\rm erg\,s^{-1}}$, provided that the radioactive decay energy is efficiently trapped near maximum light \citep{Arnett1982,Nadyozhin1994}. For an ejecta mass of $\sim1$--$10\,\msun$, a characteristic expansion speed of $v\sim0.1\,c$, and an assumed effective opacity of $\kappa\sim0.05$--$0.1\,{\rm cm^2\,g^{-1}}$, a simple radioactive-diffusion estimate gives a peak time of roughly $1$--$3$ weeks. Such transients would generally rise on timescales comparable to the fastest hydrogen-poor superluminous SNe, whose rise times typically range from $\sim20$ to $\gtrsim100$ days \citep{GalYam2019}.

The accompanying $\sim0.1\,\msun$ of $r$-process nuclei may also leave observable signatures, even if the decay of ${}^{56}\mathrm{Ni}$ and ${}^{56}\mathrm{Co}$ dominates the radioactive power on day-to-week timescales. An early $r$-process-powered component could emerge if the enriched material reaches sufficiently low optical depths while much of the nickel-powered radiation remains trapped. Moreover, $r$-process enrichment can modify the colors and spectra of the nickel-powered transient, potentially producing an infrared excess or suppressing optical emission \citep{Barnes2022,Sarin2026}. The strength and timing of these signatures depend on the local abundance pattern and the spatial and velocity distributions of the enriched material relative to the nickel-rich ejecta.

In the strongly magnetized model, the jet breaks out of the star after $\approx 1\,\rm{s}$, and the central engine injects $E_{\rm j} \sim 10^{54}\,\rm{erg}$ into the jets during the simulated $\sim 4\,\rm{s}$. This large energy budget suggests that such stars could power rare, exceptionally energetic GRBs \citep[see also][]{Gottlieb2026}, potentially including extreme events such as GRB~221009A \citep{Burns2023}. However, allowing the BH mass and spin to evolve could substantially modify the jet power and integrated energy \citep[see e.g.,][]{gottlieb_spinning_2025}.

\subsection{$r$-process yields} \label{sec:rprocess}

Our direct-horizon collapsar models produce $M_{\rm r-p}\sim0.1$--$0.3\,\msun$ of $r$-process elements, but only $M_{\rm La} \sim 10^{-5}$-- $10^{-4}\,\msun$ of lanthanides. Neglecting neutrino irradiation after the ejecta cools through $T_0$ raises $M_{\rm La}$ by a factor of $\lesssim 3$. For comparison, the kilonova of GW170817 implies $M_{\rm La} \sim 10^{-4}$--$10^{-3}\,\msun$, from a total ejecta mass of $\approx 0.05$--$0.08\,\msun$ \citep{Kasen2017,Villar2017} and a lanthanide mass fraction of $X_{\rm La} \sim 10^{-2.2 \pm 0.5}$ \citep{Ji2019}. Thus, the heavy $r$-process yield of a direct-horizon collapsar might be comparable to that of a single BNS merger. This picture differs from the semi-analytic predictions of \citet{siegel_super-kilonovae_2022}, in which disk outflows from collapsars above the pair-instability mass gap eject $\gtrsim5$--$10\,\msun$ of $r$-process elements and only $\sim 0.1$--$1\,\msun$ of ${}^{56}\rm Ni$, powering lanthanide-rich ``super-kilonovae''. Our disks indeed become neutron-rich, but the traced ejecta are $\approx45\%\,{}^{56}\rm Ni$ by mass, powering a ${}^{56}\rm Ni$-decay transient rather than a red super-kilonova.

Ordinary collapsars face the same tension between disk neutronization and mass ejection. Unbinding substantial amounts of material requires strongly magnetized disks at high accretion rates \citep{Issa2025}, yet magnetized disks accreting at $\dot{M} \lesssim 0.4\,\msun\,\rm{s}^{-1}$ cannot neutronize below $\ye \approx 0.33$ \citep{Gottlieb2026}. The extreme accretion rates of direct-horizon collapsars allow their disks to reach $\ye \approx 0.1$. However, the strong magnetic flux required to drive outflows also rapidly depletes the disk and lifts its electron degeneracy, limiting the neutron-rich ejecta, $\ye < 0.25$, to gas that neutronizes during a brief, sub-second window.

\subsection{Contribution to $r$-process enrichment at low metallicities} \label{sec:enrichment}

Direct-horizon collapsars eject both iron-group and $r$-process nuclei in a single event. If the Eu- and Fe-bearing ejecta are retained and mixed in the same proportions, their yield ratio sets the abundance ratio contributed by the event. In model $\Ps$, the ejecta contain $M_{\rm Eu} \approx 1.9\times10^{-6}\,\msun$ of europium ($Z=63$), computed from the total yields at $A = 151$ and $153$, which decay to the two stable Eu isotopes, and $M_{\rm Fe} \approx 3.6\,\msun$ of iron ($A = 56$; Fig.~\ref{fig:yields}), i.e., $M_{\rm Eu}/M_{\rm Fe} \approx 5\times10^{-7}$, which corresponds to $[{\rm Eu/Fe}] \approx +0.3$ relative to the solar abundances\footnote{The abundance ratio of elements $A$ and $B$ relative to the Solar ratio is defined as $[\rm{A/B}] \equiv \log_{10}\left(N_{\rm A}/N_{\rm B}\right) - \log_{10}\left(N_{A}/N_{\rm B}\right)_{\odot}$.} \citep{Asplund2009}. In model $\Pw$, $M_{\rm Eu} \approx 8\times10^{-8}\,\msun$ and $M_{\rm Fe} \approx 0.75\,\msun$, i.e., $[{\rm Eu/Fe}] \approx -0.4$.\footnote{Varying the $\sigma_{\max}$ threshold that defines the initial relaxation shock by $\pm0.5\,$dex (Appendix~\ref{appendix:tracers_comp}) changes $M_{\rm Fe}$ but not $M_{\rm Eu}$, giving $[{\rm Eu/Fe}] \approx 0.15$--$0.5$ in $\Ps$ and $\approx -0.6$ to $-0.2$ in $\Pw$.} Model $\Ps$ thus may reach $[{\rm Eu/Fe}] \approx 0.3$, the lower threshold of the $r$-process-enhanced stars \citep[r-I stars;][]{Holmbeck2020}. It remains to be seen how these ratios change once the BH growth, self-gravity, and nuclear recombination are accounted for.

The iron yield and subsequent dilution together determine the iron abundance of the enriched gas. Adopting the spherical-expansion and homogeneous-mixing estimate of \citet{Magg2020}, an explosion with kinetic energy $E_{\rm kin}$ in a uniform, ionized medium has a minimum dilution mass
$M_{\rm dil,min} \approx 1.9\times10^{4}\,\msun\,(E_{\rm kin}/10^{51}\,{\rm erg})^{0.96}\,(n_0/{\rm cm^{-3}})^{-0.11}$,
where $n_0$ is the ambient number density. For negligible pre-existing iron, this gives
$[{\rm Fe/H}] \lesssim \log_{10}[M_{\rm Fe}/(X_{\rm Fe,\odot}M_{\rm dil,min})]$,
where $X_{\rm Fe,\odot} \approx 1.3\times10^{-3}$ is the solar iron mass fraction \citep{Asplund2009}. Taking $M_{\rm Fe} \approx 0.45\,M_{\rm ej}$ and a characteristic ejecta speed $v \sim 0.1\,c$, the kinetic energy is
$E_{\rm kin} \approx M_{\rm ej}v^2/2 \approx 9\times10^{51}\,{\rm erg}\,(M_{\rm ej}/\msun)$.
For $n_0 \sim 1\,{\rm cm^{-3}}$ and $M_{\rm ej}\sim10\,\msun$, this yields $[{\rm Fe/H}] \lesssim -2.6$. Since $M_{\rm Fe}/M_{\rm dil,min}\propto M_{\rm ej}^{0.04}$, the estimate is largely independent of the uncertain ejecta mass, provided that the iron fraction and characteristic velocity remain unchanged. Additional jet energy that couples to the same gas, beyond that already included in $E_{\rm kin}$, would increase the dilution mass and lower this limit further. Stronger dilution can therefore produce extremely metal-poor gas, but cannot increase its Eu/Fe ratio if the Eu- and Fe-bearing ejecta are retained and mixed in the same proportions.

If GW231123-like mergers originate from this channel, their inferred rate, $0.08^{+0.19}_{-0.07}\,\rm{Gpc^{-3}\,yr^{-1}}$ \citep{LVK2025_GW231123}, samples only the subset of the parent population whose remnants enter suitable binaries and merge within a Hubble time. The total collapse rate could be substantially higher, leaving room for an appreciable contribution even at low redshift. The case for an important contribution is particularly compelling at early times: reduced line-driven winds at low metallicity favor the retention of helium cores above the pair-instability gap \citep{Vink2001,Heger2003,Vink2005}, as in the $Z\approx0.07\,Z_\odot$ progenitor used here \citep{Renzo2020b}, and the fraction of star formation at $Z\lesssim0.1\,Z_\odot$ increases with redshift \citep{Langer2006,Chruslinska2019}. These conditions favor direct-horizon collapsar formation in the early Universe, where their short stellar lifetimes allow enrichment within a few Myr of star formation.

Taken together, the yields, dilution estimates, and potentially higher formation rates make direct-horizon collapsars plausible prompt contributors to the $r$-process enrichment of individual low-metallicity systems. The principal nucleosynthetic tension in our models is the modest ${\rm [Eu/Fe]}$ compared with strongly enhanced systems such as Reticulum~II \citep{Ji2016_nature,Ji2016}, rather than the absence of $r$-process production or the inability to enrich gas at low ${\rm [Fe/H]}$. This ratio is particularly sensitive to the uncertain iron yield; a reduction in the iron yield without a comparable reduction in the Eu yield would raise ${\rm [Eu/Fe]}$ and could alleviate the tension. 

\begin{acknowledgments}

We thank Mathieu Renzo, Rebecca Surman, and the participants of the ``Enigmatic Explosions'' program at the Kavli Institute for Theoretical Physics (KITP) for stimulating discussions. 
This research was supported in part by grant NSF PHY-2309135 to the KITP.
D.I. was partially supported by NSF AST-2407475.
This research used resources of the Argonne Leadership Computing Facility, a U.S. Department of Energy (DOE) Office of Science user facility at Argonne National Laboratory and is based on research supported by the U.S. DOE Office of Science Advanced Scientific Computing Research Program, under Contract No. DE-AC02-06CH11357 (NeutronStarRemnants project).
The authors acknowledge the MIT Office of Research Computing and Data for providing high performance computing resources on the Engaging cluster that have contributed to the research results reported within this paper.
We acknowledge the use of Claude Opus 5.5 (Anthropic) for coding assistance and simulated peer review. The authors reviewed all AI-generated outputs and take full responsibility for the contents of the final manuscript.
\end{acknowledgments}

\software{\texttt{$\nu$H-AMR} \citep{hamr_paper,Issa2025}, \texttt{GR1D} \citep{OConnor2010}, \texttt{NuLib} \citep{oconnor_open-source_2015}, \texttt{SkyNet} \citep{Lippuner2017}, \texttt{ParaView} \citep{Paraview_Ahrens2005}, \texttt{numpy} \citep{numpy_paper}, \texttt{scipy} \citep{scipy_paper}, \texttt{matplotlib} \citep{matplotlib_paper}}

\bibliography{refs}{}
\bibliographystyle{aasjournalv7}

\appendix

\section{Neutrino transport} \label{appendix:nu_transport}

Neutrinos are crucial in determining the composition of the gas through electron and positron captures. In \texttt{$\nu$H-AMR} we use two-moment (M1) neutrino transport following the implementation by \citet{sadowski_semi-implicit_2013,mckinney_three-dimensional_2014,foucart_post-merger_2015,foucart_impact_2016}. This assumes that radiation is isotropic in a ``radiation'' frame, also known as Levermore closure \citep{Levermore1984}, characterized by a 4-velocity $u_{\rm R}^\mu$,
\begin{equation}
\begin{split}
    T^{\mu\nu} &= (\rho + u_{\rm g} + P_{\rm g} + b^2) u^\mu u^\nu + (P_{\rm g} + b^2/2) g^{\mu\nu} - b^\mu b^\nu, \\
    R_{\nu_i}^{\mu\nu} &= \frac{E_{\rm R}}{3} \left( 4u_{\rm R}^\mu u_{\rm R}^\nu + g^{\mu\nu}\right), \\
    \nabla_\nu &\left( T^{\mu\nu} + \sum_{i} R_{\nu_i}^{\mu\nu} \right) = 0,
    \label{eq:stress-energy-tensor}
\end{split}
\end{equation}
where $b^\mu$ is the magnetic field 4-vector measured in a fluid frame, $u^\mu$ is the gas 4-velocity, and $E_{\rm R}$ is the neutrino energy density measured in the ``radiation'' frame. We evolve three species of neutrinos, $\nu_i = (\nu_{\rm e}, \bar{\nu}_{\rm e}, \nu_{\rm x})$, which are electron neutrinos, electron antineutrinos, and all of the heavy lepton neutrinos, respectively. We evolve the flux and geometric source terms explicitly and integrate the neutrino--matter source terms implicitly. This scheme is accurate in our disks, where the optical depths to neutrino absorption and scattering remain $\lesssim$ a few. The implementation details of the neutrino transport scheme, including the evolution of $\ye$, and a suite of tests are described in \citet{Issa2025}.

We tabulate neutrino emissivities and opacities with \texttt{NuLib} \citep{oconnor_open-source_2015} for three species: $\nu_{\rm e}$, $\bar{\nu}_{\rm e}$, and $\nu_x$ (representing $\nu_{\mu}$, $\bar{\nu}_{\mu}$, $\nu_{\tau}$ and $\bar{\nu}_{\tau}$). The interactions that we include are 1) $\nu_{\rm e}$ absorption on neutrons, 2) $\bar{\nu}_{\rm e}$ absorption on protons, and 3) production of neutrino--antineutrino pairs of all flavors from $e^{+}e^{-}$ annihilation and nucleon--nucleon bremsstrahlung. All inverse reactions are also included, in such a way that the emissivity $\eta$ and absorption opacity $\kappa_{\rm a}$ satisfy Kirchhoff's law $\eta/\kappa_{\rm a} = B_{\nu}$, with $B_{\nu}$ the black-body distribution function of neutrinos in equilibrium with the fluid, integrated over the relevant energy bin. We also include elastic scattering on protons and neutrons. The table assumes a composition of free nucleons, and the spectral quantities are averaged over energy to give gray emissivities and opacities. When post-processing the tracers, we correct two approximations in the resulting $\ye$ source term: the rescaling of the absorption opacities to the neutrino spectrum, and the inclusion of pair processes in the $\nu_e$ and $\bar\nu_e$ rates (Appendix~\ref{appendix:tracers_ye}).

\section{Resolving the MRI in the disk} \label{appendix:mri_criterion}
\setcounter{equation}{0}

\begin{figure}
    \centering
    \includegraphics[width=\columnwidth]{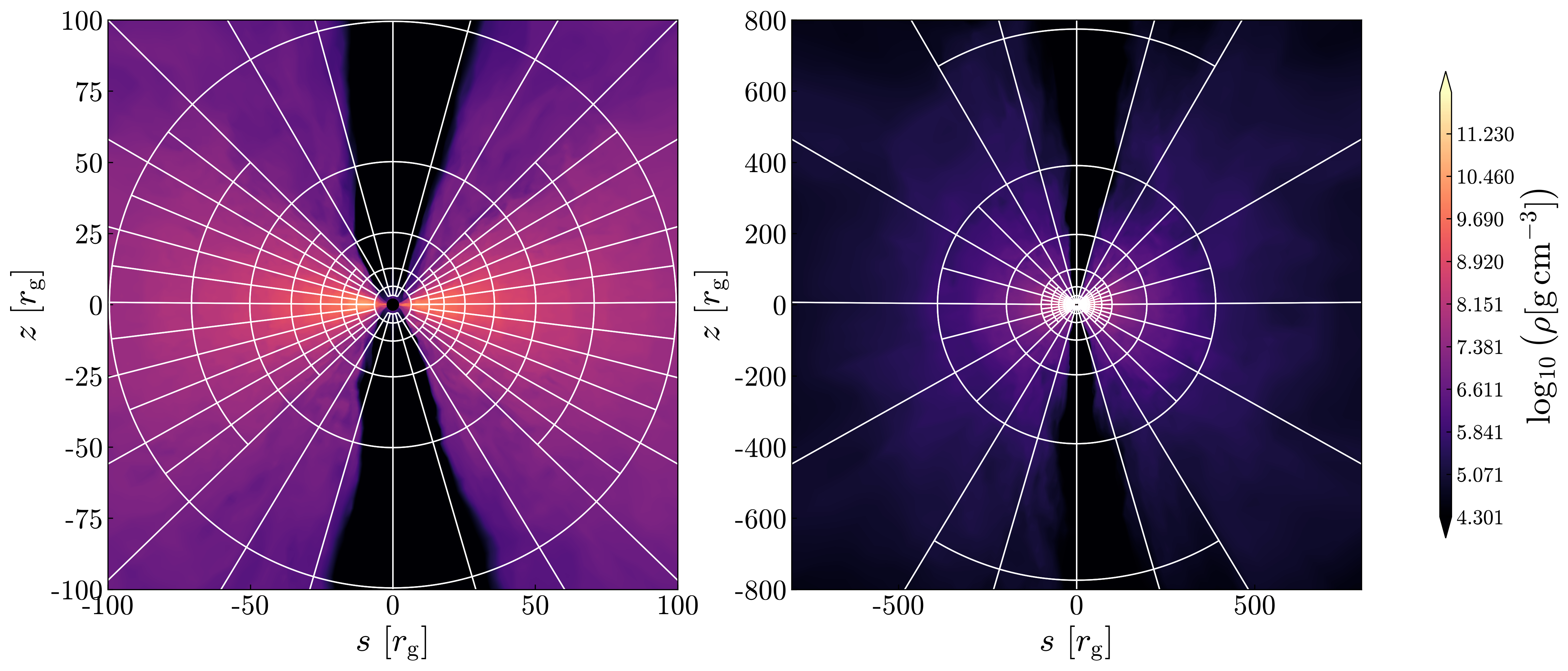}
    \includegraphics[width=0.6\columnwidth]{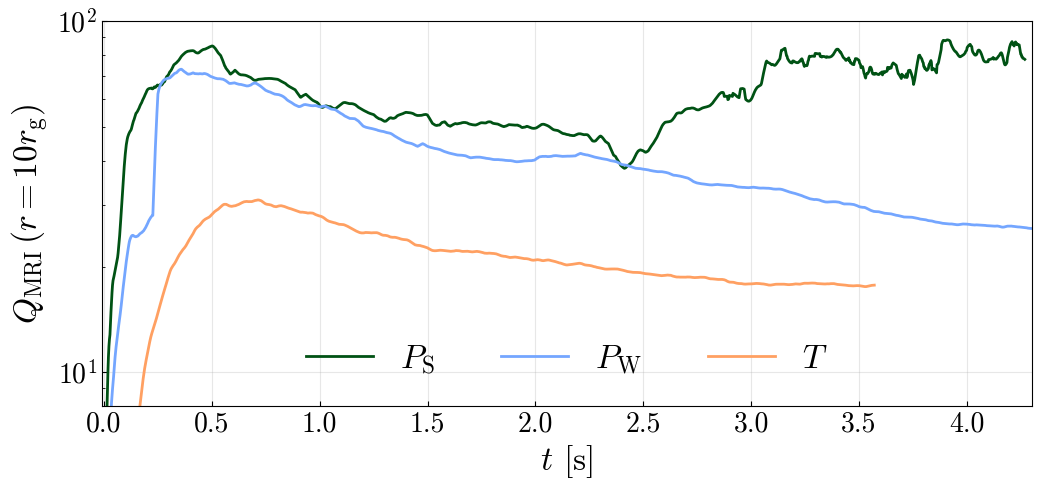}
    \caption{Top: meridional slices of the logarithm of density in model $\Pw$ at $t = 4.29\,\rm{s}$, within $100\,\rg$ (left) and $800\,\rg$ (right). White lines mark the mesh-block boundaries, showing the SMR structure. Bottom: MRI quality factor, $Q_{\rm MRI}$, at $r = 10\,\rg$ as a function of time in the magnetized models.}
    \label{fig:MRI_disk}
\end{figure}

Fig.~\ref{fig:MRI_disk} (top) shows the SMR grid in model $\Pw$, with the two refinement layers concentrated in the disk. We quantify how well the grid resolves the MRI through the quality factor \citep{hawley_assessing_2011},
\begin{equation}\label{eq:QMRI}
    Q_{\rm MRI} = \frac{\lambda_{\rm MRI}}{\Delta x^{\hat\theta}}, \qquad \lambda_{\rm MRI} = \frac{2\pi\, |b^{\hat\theta}|}{\Omega \sqrt{\rho h + b^2}},
\end{equation}
where $\lambda_{\rm MRI}$ is the wavelength of the fastest-growing MRI mode, $b^{\hat\theta}$ and $\Delta x^{\hat\theta}$ are the polar component of the magnetic field and the cell size in the fluid frame, and $h$ is the specific enthalpy. The MRI is adequately resolved for $Q_{\rm MRI} \gtrsim 10$ \citep{hawley_assessing_2011}. Fig.~\ref{fig:MRI_disk} (bottom) shows that the density-weighted disk average of $Q_{\rm MRI}$ at $r = 10\,\rg$ exceeds this threshold by $t \approx 0.1\,{\rm s}$ in models $\Ps$ and $\Pw$ and by $t \approx 0.15\,{\rm s}$ in model $\Ts$, and remains above it for the rest of the simulations.

\section{Tracer post-processing} \label{appendix:tracers}

Computing $r$-process yields requires the thermodynamic history of the ejecta. Since nuclear reaction networks are too expensive to evolve on the fly in global simulations, we follow the standard approach of post-processing Lagrangian tracer trajectories with a network. This neglects the effect of $r$-process heating on the ejecta dynamics. The procedure below aims at yields accurate to within an order of magnitude, comparable to the uncertainty from the dynamics (Section~\ref{sec:caveats}); we do not attempt to reduce the post-processing errors below this level.

\subsection{Trajectory reconstruction} \label{appendix:tracers_traj}

\begin{figure}
    \centering
    \includegraphics[width=\columnwidth]{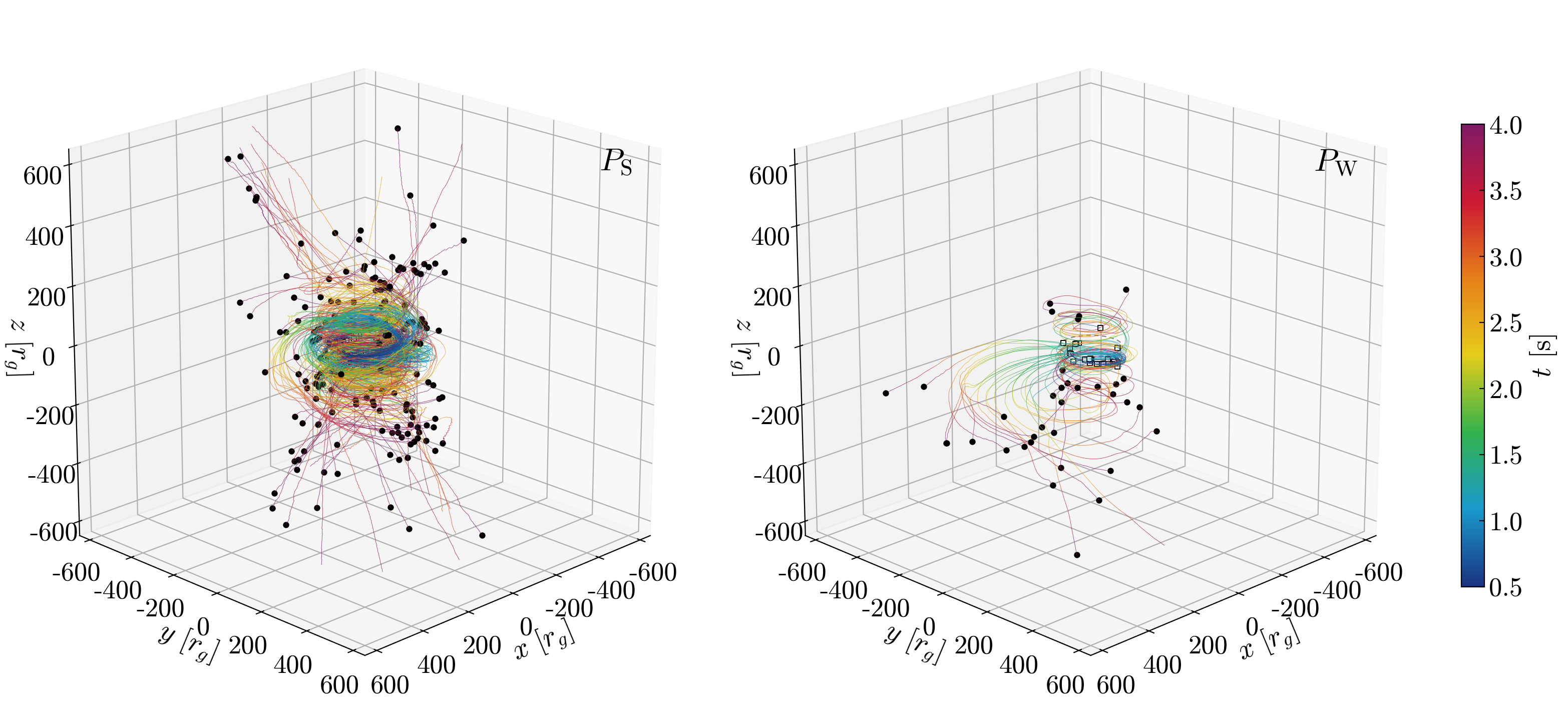}
    \caption{Trajectories of the tracers with $\ye(T_0)\leq0.25$ in models $\Ps$ (left) and $\Pw$ (right), colored by time. Unfilled squares mark the start of the \texttt{SkyNet} trajectories, where the tracers last cool through $T_0$, close to the disk (dashed gray circle marks $r = 100\,\rg$ in the equatorial plane). Black dots mark the tracer locations at the end of the simulation. In $\Ps$, the tracers circulate around the disk at $|z| \lesssim 200\,\rg$ until $t \approx 2.5\,\rm{s}$ and then escape along the outflows to $|z| \sim 100$--$600\,\rg$. In $\Pw$, they cool through $T_0$ at $t \approx 1\,\rm{s}$ and then spread to $r \approx 200$--$600\,\rg$, mostly below the equatorial plane.}
    \label{fig:tracer_trajectories}
\end{figure}

Our \texttt{$\nu$H-AMR} simulations were run without tracer particles, so we reconstruct the trajectories from the simulation data (Fig.~\ref{fig:tracer_trajectories}), output every $\Delta t_{\rm dump}=10^2\,\rg/c\approx12\,\rm{ms}$.\footnote{For models $\Pw$ and $\Ts$, the simulation data are coarsened by a factor of $2$ per dimension before tracing. We verified on model $\Ps$ that tracing on the reduced data, coarsened by a factor of $4$ per dimension and output at $20\,\rg/c$ cadence, yields the same results (for $r_{\rm ext}=300\,\rg$).} We sample $\mathcal{N}_{\rm T}=3\times 10^5$ tracers from the unbound ($\mathcal{B}>0$), outflowing ($u^r>0$) gas, $10^5$ in each of three populations: gas crossing the extraction sphere $r_{\rm ext}=10^{3}\,\rg$ during the first and the second half of the simulation, sampled in proportion to the crossing mass, and gas at $20\,\rg\leq r<r_{\rm ext}$ at the last dump, sampled in proportion to the cell mass. Each tracer carries an equal share of its population's mass. This oversamples the late and interior ejecta, which carry most of the low-$\ye$ gas, without biasing the mass-weighted yields.

We integrate each tracer backward from its seeding time to $t=0$ and, for the crossing populations, also forward to the last dump. We use the coordinate velocity $v^i=u^i/u^t$, interpolated trilinearly in the code coordinates and linearly in time, with second-order Runge--Kutta (midpoint) substeps; the number of substeps per dump interval is a power of two, chosen for each tracer to keep the Courant number in the code coordinates below $0.05$. At every dump we record $\rho$, $T$, $\ye$, $\mathcal{B}$, the magnetization $\sigma\equiv b^2/\rho$, the specific angular momentum $u_\varphi$, and the fluid-frame number densities and mean energies of $\nu_e$ and $\bar\nu_e$.

\subsection{Electron fraction} \label{appendix:tracers_ye}

Along each trajectory we re-integrate $\ye$ from $t=0$, keeping $\rho$, $T$ and the neutrino radiation field as simulated, but correcting two approximations in the $\ye$ source term of the simulations. First, the absorption opacities, tabulated for neutrinos in equilibrium with the gas, were rescaled to the actual neutrino spectrum by $(T_\nu/T)^2$, where $T_{\nu}$ is the neutrino temperature \citep{foucart_impact_2016}. We instead rescale them by the ratio of the spectrum-averaged charged-current cross sections, $\sigma_{\rm cc}\propto E_e\sqrt{E_e^2-m_e^2}$ with $E_e=E_\nu\pm\Delta$ and $\Delta=m_n-m_p$, which includes the $\bar\nu_e$ capture threshold. For the neutrino spectrum, we take a Fermi--Dirac distribution with the degeneracy of $\beta$-equilibrium with the gas and a temperature matching the simulated mean energy. Second, the \texttt{NuLib} $\nu_e$ and $\bar\nu_e$ emissivities and absorption opacities include pair annihilation and nucleon--nucleon bremsstrahlung, which the $\ye$ source counted as captures on nucleons; we instead take them from a \texttt{NuLib} table computed on the same grid without these processes. Since \texttt{NuLib}'s gray $\bar\nu_e$ opacity is unreliable at $T\lesssim 1\,$MeV, we read it at $\max(T,1.5\,{\rm MeV})$ and convert it to the gas temperature with the same cross-section ratio. The resulting number opacities agree with \texttt{NuLib}'s spectral charged-current opacities within $\approx15\%$ for $10^5\lesssim\rho\lesssim10^{10}\,{\rm g\,cm^{-3}}$ and $0.3\lesssim T\lesssim2\,$MeV. Between consecutive dumps we average the rates over the two ends of the interval and integrate the resulting linear relaxation of $\ye$. With the rates used in the simulations, this procedure reproduces the simulated $\ye$ at seeding (median $|\Delta\ye|\lesssim0.003$). 

\subsection{Nucleosynthesis} \label{appendix:tracers_nuc}

We start the nuclear reaction network \texttt{SkyNet} \citep{Lippuner2017} in nuclear statistical equilibrium (NSE) when each tracer last cools through $T_0 = \min(T_{\rm tr;\,peak}, 8\,{\rm GK})$. Tracers with $5\,{\rm GK}\leq T_{\rm tr;\, peak}<8\,$GK ($19$--$25\%$ of the mass, depending on the model) thus start in NSE at their peak, with the $\ye$ there. The few tracers still hotter than $8\,$GK at the end of their trajectory ($\lesssim0.4\%$ of the ejecta mass) start there. We discard the tracers with $T_{\rm tr;\, peak}<5\,$GK ($40$--$62\%$ of the mass): since our pipeline does not track the stellar composition and sets the initial $\ye=0.5$, starting them in NSE would turn them almost entirely into $A=56$, instead of retaining their original composition (O-Ne-Mg, C below $2\,\rm{GK}$) or burning explosively.

\texttt{SkyNet} evolves the composition with a prescribed density and computes the temperature. The density follows the simulation to the end of the trajectory and then the homologous continuation $\rho\propto t^{-3}$, with $t$ measured from the start of the simulation, blended in over the last $20\%$ of the trajectory. The temperature starts from the tracer's temperature at the NSE start and evolves through expansion and nuclear self-heating, so heating by later shocks is not included. We use the Helmholtz EoS \citep{timmes_accuracy_2000}, strong and weak reaction rates from REACLIB \citep{Cyburt2010}, neutron-induced fission with symmetric fragments and no prompt neutrons \citep{Panov2010}, spontaneous fission from the approximation of \citet{Frankel1947} with the barriers of \citet{Mamdouh2001}, the nuclear masses distributed with REACLIB, measured (AME2011) where available and from the finite-range droplet macroscopic model \citep[FRDM; see, e.g.,][]{Moller2016} otherwise, partition functions from \citet{Rauscher2000}, and the screening distributed with \texttt{SkyNet} \citep{Lippuner2017}. We evolve each trajectory to $t = 10^9\,{\rm s}$.

We run each tracer twice: with the charged-current reactions on free nucleons distributed with \texttt{SkyNet}, $p+e^-\leftrightarrow n+\nu_e$ and $n+e^+\leftrightarrow p+\bar\nu_e$ \citep[][App.~C, with the matrix element and normalization of \citealt{Arcones2010}]{Lippuner2017}, and without them. Below $T_0 \lesssim 8\,\rm{GK}$, $e^{\pm}$ captures are negligible compared with $\nu$ absorptions near the disk, so the two post-processing runs differ mainly by the absorptions.
In the first run, the $\nu_e$ and $\bar\nu_e$ captures follow the tracer's own neutrino history: a Fermi--Dirac spectrum with zero chemical potential, with the temperature set by the mean energy and the occupation normalized to the number density, diluted as $t^{-2}$ after the simulation ends. The captures do not heat the gas. To reduce the cost, we evolve all tracers with $\ye<0.3$ at the NSE start, which produce the heavy $r$-process elements ($1086$ in $\Ps$, $154$ in $\Pw$, and none in the models without jets), and random subsets of $3\times10^{3}$ of the remaining tracers in each of the two sets (starting at $8\,$GK and at the peak), with their masses rescaled to conserve the mass of each set. The heavy-element yields are dominated by $N_{\rm eff}\approx20$ ($\Ps$) and $10$ ($\Pw$) tracers, implying a sampling uncertainty of $\approx0.1\,$dex. The yields are the mass-weighted sums over all tracers.

\subsection{Ejecta components} \label{appendix:tracers_comp}

\begin{figure*}
    \centering
    \includegraphics[width=\linewidth]{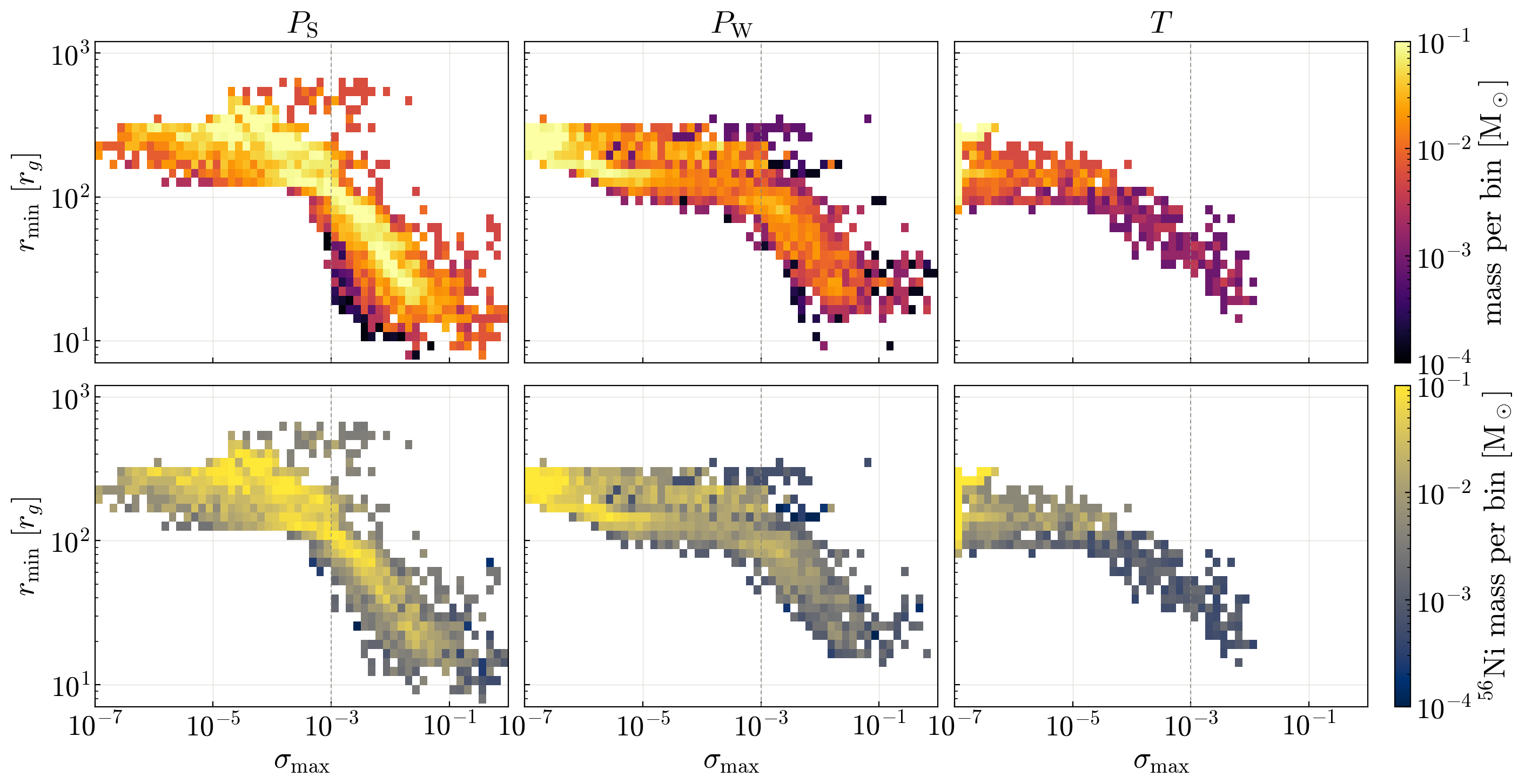}
    \caption{Mass (top) and ${}^{56}$Ni mass (bottom) of the traced unbound ejecta with $T_{\rm tr;\,peak}\geq5\,$GK, including all tracer classes, as a function of the maximum magnetization along the trajectory, $\sigma_{\max}$, and the minimum radius, $r_{\rm min}$, in models $\Ps$, $\Pw$, and $\Ts$. Vertical dashed lines mark the cocoon threshold, $\sigma_{\max} = 10^{-3}$.}
    \label{fig:rmin_sigmamax}
\end{figure*}

\begin{figure}
    \centering
    \includegraphics[width=\columnwidth]{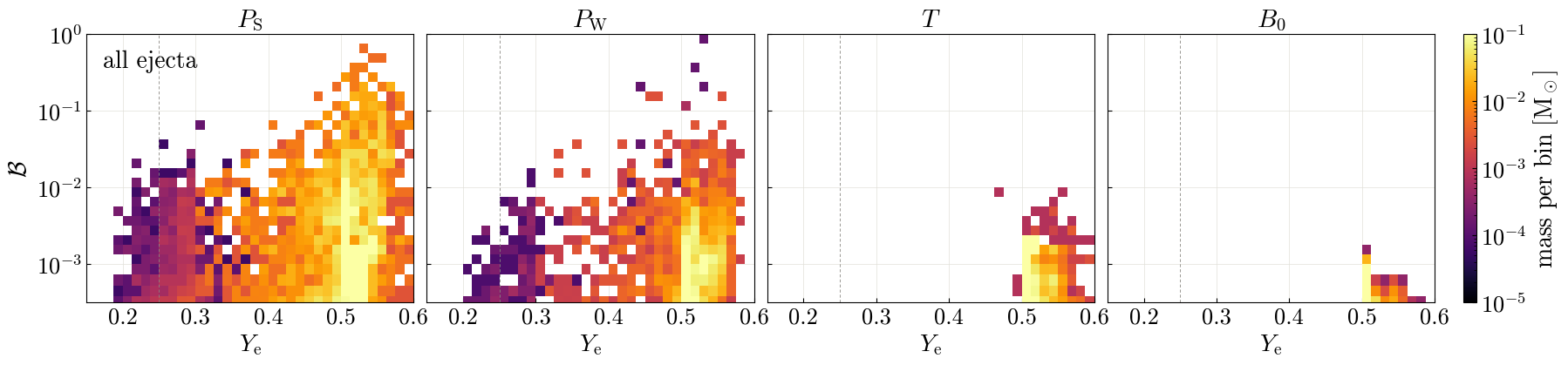}
    \includegraphics[width=\columnwidth]{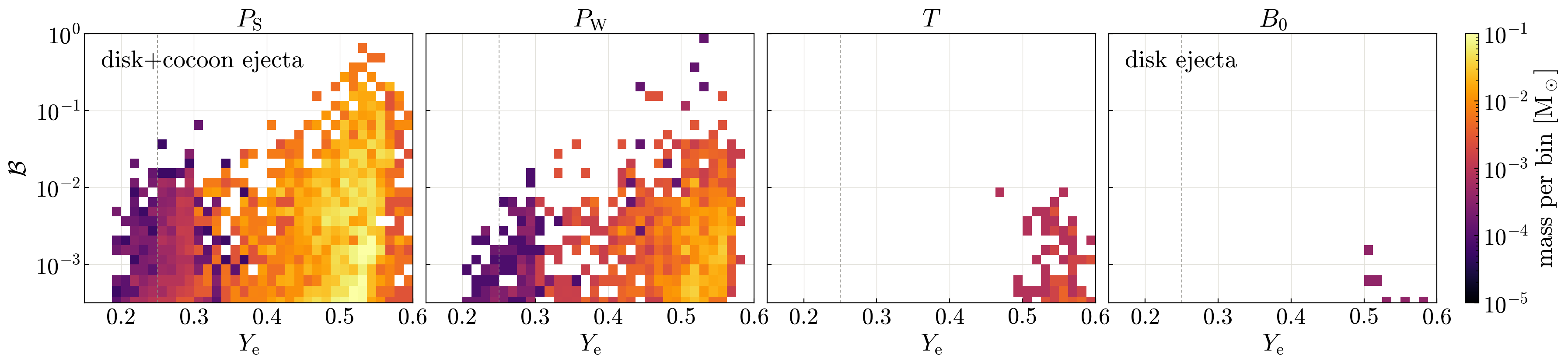}
    \caption{Mass distribution of the traced unbound ejecta in $\ye$ and the Bernoulli parameter, $\mathcal{B}$ [Eq.~\eqref{eq:bernoulli_mhd}], in all models. Top row: all traced ejecta. Bottom row: disk and cocoon ejecta (disk ejecta only in $\Bz$), i.e., excluding the shock-heated stellar gas, which is the subset shown in Figs.~\ref{fig:Ye_tracers} and \ref{fig:yields}. Vertical dashed lines mark $\ye = 0.25$. Most of the ejecta are marginally unbound, $\mathcal{B} \lesssim 10^{-2}$, and concentrated at $\ye \approx 0.45$--$0.55$. Only models $\Ps$ and $\Pw$ eject material with $\ye < 0.25$, all of it with $\mathcal{B} \lesssim 0.03$. In $\Ts$ and $\Bz$, nearly all of the ejecta at $\ye \approx 0.5$ is shock-heated stellar gas. Only tracers with $T_{\rm tr;\,peak} \geq 5\,$GK are shown; $\ye$ is measured at the peak temperature, and $\mathcal{B}$ at the end of the simulation.}
    \label{fig:ejecta_ye_B}
\end{figure}

\begin{figure}
    \centering
    \includegraphics[width=\columnwidth]{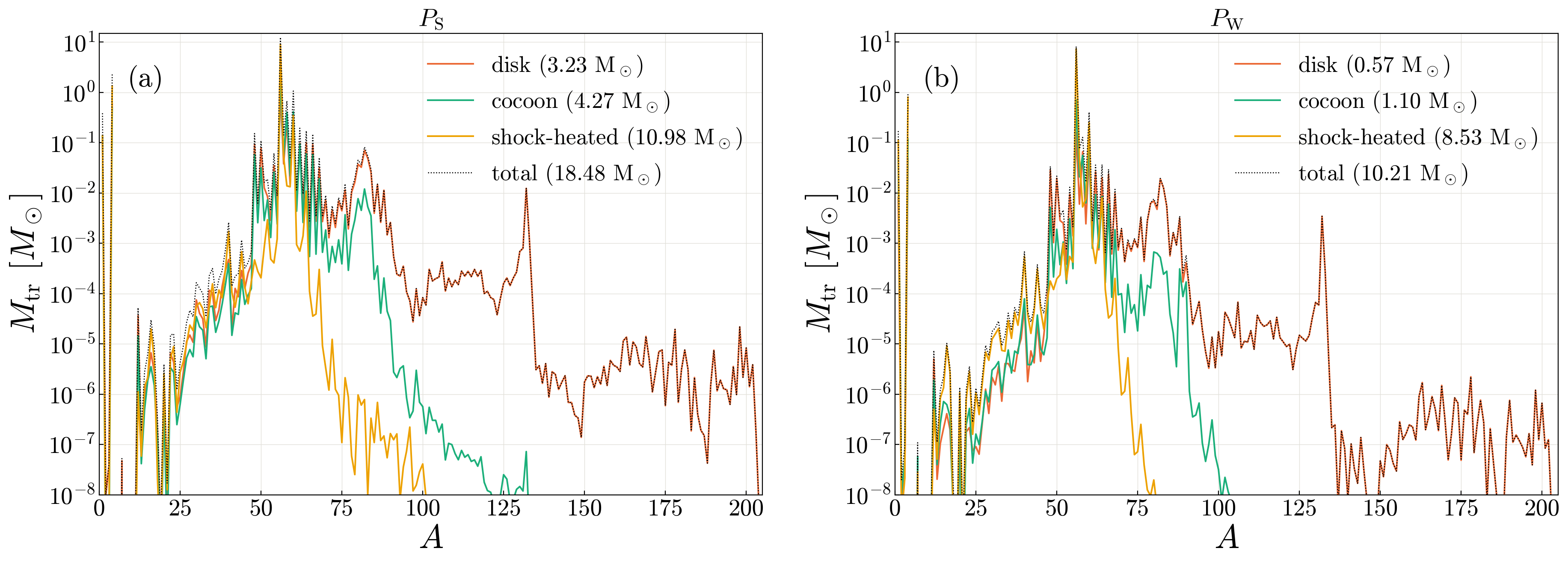}
    \caption{Nucleosynthesis yields of the traced unbound ejecta in models $\Ps$ (a) and $\Pw$ (b), split by origin into disk, cocoon, and shock-heated stellar gas, with their total masses indicated in the legend; dotted lines show the totals. Only tracers with $T_{\rm tr;\,peak}\geq5\,$GK are included. In both models, the shock-heated gas produces mostly $A = 56$ and contributes little beyond $A \approx 75$, and the disk ejecta produce nearly all nuclei with $A \gtrsim 90$, including the second and third $r$-process peaks. The cocoon ejecta extend to $A \approx 125$ in $\Ps$ and $A \approx 100$ in $\Pw$.}
    \label{fig:yields_breakdown}
\end{figure}

We classify the tracers exclusively in this order: jet tracers if they reach $\mathcal{B} > 1$; disk tracers if they spend at least $\approx 200\,\rg/c$ at $r<10^3\,\rg$ with $u_\varphi \geq 0.75\,l_{\rm K}$, where $l_{\rm K}$ is the Keplerian specific angular momentum; and cocoon tracers if $\sigma_{\max} \geq 10^{-3}$ along the trajectory. Fig.~\ref{fig:rmin_sigmamax} motivates this threshold. In the jet-launching models, the tracers form two branches in the plane of $\sigma_{\max}$ and the minimum radius along the trajectory, $r_{\rm min}$: weakly magnetized stellar gas, $\sigma_{\max} \lesssim 10^{-3}$, which never comes closer than $r_{\rm min} \approx 100\,\rg$ to the BH, and gas that does, whose $\sigma_{\max}$ increases as $r_{\rm min}$ decreases. The two branches meet at $\sigma_{\max} \approx 10^{-3}$, most clearly in $\Ps$. In $\Ts$, the distribution is continuous, and the gas with $\sigma_{\max} \geq 10^{-3}$ carries little mass. Varying the threshold by $\pm0.5\,$dex, to $\sigma_{\max} = 10^{-3.5}$ ($10^{-2.5}$), changes the disk and cocoon ejecta to $\approx 9.4$ ($5.8$) and $2.1$ ($1.1$)$\,\msun$ in $\Ps$ ($\Pw$), and their iron mass to $\approx 5.0$ ($2.4$) and $1.3$ ($0.5$)$\,\msun$. Since the $r$-process nuclei and lanthanides originate almost entirely in the disk ejecta, their yields remain unchanged. The remaining tracers are shock-heated stellar gas: the relaxation of the remapped initial conditions drives a shock that burns the outer layers of the star into excess iron. Since this is an artifact of the remapping, we exclude these tracers from the analysis; most of them have $T_{\rm tr;\,peak}<5\,$GK ($53$--$62\%$ of their mass). We also exclude the jet tracers, $\approx 1\%$ of the traced mass in $\Ps$ and $0.3\%$ in $\Pw$, whose thermodynamics is affected by the density and internal energy floors. The disk and cocoon ejecta amount to $\approx 7.5\,\msun$ in $\Ps$ and $\approx 1.7\,\msun$ in $\Pw$ (Fig.~\ref{fig:Ye_tracers}), about $24\%$ and $7\%$ of the traced mass ($31.2$ and $23.7\,\msun$); in the models without jets, almost all tracers are shock-heated. Fig.~\ref{fig:ejecta_ye_B} compares the $\ye$ and $\mathcal{B}$ distributions of all traced ejecta and of the disk and cocoon ejecta. Fig.~\ref{fig:yields_breakdown} shows the yields of each component in $\Ps$ and $\Pw$, from the tracers with $T_{\rm tr;\,peak}\geq5\,$GK: in both models, the disk ejecta produce nearly all nuclei with $A \gtrsim 90$ and the shock-heated ejecta produce mostly $A = 56$, while the cocoon ejecta extend to $A \approx 125$ in $\Ps$ and $A \approx 100$ in $\Pw$.

\end{document}